\documentclass[10pt,a4paper]{article}

\usepackage[utf8]{inputenc}
\usepackage[T1]{fontenc}
\usepackage[english]{babel}

\usepackage{amsmath,amssymb,amsthm}
\usepackage{graphicx}
\usepackage{xcolor}
\usepackage{placeins}
\usepackage{booktabs}
\usepackage{siunitx}
\usepackage{authblk}
\usepackage{hyperref}

\newtheorem{Remark}{Remark}

\providecommand{\keywords}[1]{%
  \par\smallskip
  \noindent\textbf{Keywords:} #1
}

\title{Singular barriers and quartic integrability breaking in the TTW system}

\author[1]{Adrian M. Escobar Ruiz\thanks{\texttt{admau@xanum.uam.mx}}}
\author[1]{Miguel E. G\'omez Quintanar\thanks{\texttt{cbi2213009326@izt.uam.mx}}}
\author[1]{Lidia Jim\'enez-Lara\thanks{\texttt{lidia@xanum.uam.mx}}}
\author[2,3]{Jaume Llibre\thanks{\texttt{jaume.llibre@uab.cat}}}

\affil[1]{%
Departamento de F\'isica, Universidad Aut\'onoma Metropolitana Unidad Iztapalapa,\\
San Rafael Atlixco 186, 09340 Ciudad de M\'exico, M\'exico
}

\affil[2]{%
Departament de Matem\`atiques, Universitat Aut\`onoma de Barcelona,\\
08193 Bellaterra, Barcelona, Catalonia, Spain
}

\affil[3]{%
Reial Acad\`emia de Ci\`encies i Arts de Barcelona,\\
La Rambla 115, 08002 Barcelona, Catalonia, Spain
}

\date{}

\begin{document}

\maketitle

\begin{abstract}
We study the competition between resonance-induced integrability breaking and centrifugal confinement in a symmetric quartic deformation of the classical $k=1$ Tremblay--Turbiner--Winternitz (TTW) system, $H=\frac{1}{2}\left(P_X^2+P_Y^2\right)+X^2+Y^2 +\frac{\gamma}{X^2}+\frac{\gamma}{Y^2}+\kappa X^2Y^2.$ The inverse-square terms are the centrifugal remnants of an $SO(2)\times SO(2)$ reduction of a four-dimensional oscillator, while the $X^2Y^2$ interaction couples the reduced radial modes and connects the Smorodinsky--Winternitz and Contopoulos limits. For $\gamma>0$, the axes are impenetrable and split configuration space into invariant sectors. For sufficiently small nonzero $\kappa$, resonant averaging constructs a phase-locked periodic orbit on each energy surface $H=h>4\sqrt{\gamma}$ whose fixed-energy reduced Poincar\'e map has no unit characteristic multiplier. Poincar\'e's criterion therefore excludes a second independent $C^1$ first integral in any invariant neighbourhood of this orbit. At finite coupling, Poincar\'e sections and finite-time Lyapunov maps show the breakup of invariant curves and the growth of chaotic layers. Comparison with the barrier-free limit separates two effects: the quartic interaction breaks integrability, whereas the singular barriers reduce phase-space connectivity and chaotic transport without restoring it.
\end{abstract}


\section{Introduction}

The transition from integrability to chaos in Hamiltonian systems is often controlled by a competition between resonant coupling and geometric confinement. Superintegrable systems provide an especially sharp starting point for this problem: their bounded trajectories are closed, their phase space is maximally foliated by invariant structures, and small perturbations can be used to probe how this exceptional organization is destroyed \cite{Evans1990,MillerPostWinternitz2013}. {The Tremblay--Turbiner--Winternitz (TTW) system,
\[
H_{TTW} =
\frac{1}{2\,m}\left(p_r^2+\frac{1}{r^2}\,p_\theta^2\right)
+ m\,\omega^2\,r^2
+ \frac{\alpha\,k^2}{r^2\,\cos^2(k\,\theta)}
+ \frac{\beta\,k^2}{r^2\,\sin^2(k\,\theta)} \ ,
\]
where $(r,\theta)$ are polar coordinates on the plane and $\alpha$ and $\beta$ are independent barrier strengths, is a canonical example of this class \cite{FrisMandrosovSmorodinskyUhlirWinternitz1965,Evans1991,
TremblayTurbinerWinternitz2009}. For $k=1$, it reduces to the planar Smorodinsky--Winternitz system. In the present work, we restrict to the exchange-symmetric subfamily $\alpha=\beta=g>0$. This specialization combines an isotropic oscillator with equal inverse-square barriers and
admits more independent first integrals than degrees of freedom. It is therefore maximally superintegrable, and all its bounded classical trajectories are periodic \cite{TremblayTurbinerWinternitz2010,MillerPostWinternitz2013}.}

A natural way to break this structure is to add the symmetric quartic interaction $x^2y^2=r^4\,\cos^2(\theta)\,\sin^2(\theta)$.  This perturbation preserves the discrete reflection
and exchange symmetries of the $k=1$ TTW Hamiltonian, but it couples the two separable degrees of freedom. The resulting dimensionless Hamiltonian, in Cartesian-like coordinates $(X,Y)$, is of the form
\begin{equation}
{\cal H}=
\frac12(P_X^2+P_Y^2)
+X^2+Y^2
+\frac{\gamma}{X^2}
+\frac{\gamma}{Y^2}
+\kappa \,X^2Y^2 .
\label{eq:intro-H}
\end{equation}
{
The identical coefficients of the two inverse-square terms define the exchange-symmetric subfamily of the $k=1$ TTW system. The general case $\gamma_X/X^2+\gamma_Y/Y^2$ is not considered here. The choice $\gamma_X=\gamma_Y=\gamma$ is used essentially in Sec.~4 to identify the phase-locked branch $r=s$ and to establish the nondegeneracy condition entering Proposition~1. We make no claim here for arbitrary unequal barrier strengths. The inverse-square walls have a standard higher-dimensional origin. Let $X=\lVert\mathbf q_1\rVert$ and $Y=\lVert\mathbf q_2\rVert$ be the radial coordinates of two planar isotropic oscillators. Fixing their angular momenta and reducing by $SO(2)\times SO(2)$ generates the centrifugal terms $\gamma_X/X^2+\gamma_Y/Y^2$ \cite{EvansVerrier2008,RodriguezTempestaWinternitz2008}. Under this lift, the deformation becomes $\kappa\lVert\mathbf q_1\rVert^2\lVert\mathbf q_2\rVert^2$: it preserves both planar rotations while coupling the radial modes. The construction makes the loss of the familiar oscillator integrals plausible, but it does not exclude an additional hidden invariant of the reduced two-degree-of-freedom problem. Indeed, the coupled four-dimensional parent system retains the Hamiltonian and the two planar angular momenta, whereas Liouville integrability would require a fourth commuting invariant. The analytic purpose of the present work is to provide the missing obstruction through a continued resonant orbit with a non-unit characteristic multiplier. The numerical analysis addresses a separate question: how strongly can centrifugal barriers suppress chaotic transport after integrability has been lost? The need for a dynamical obstruction is illustrated by the quartic H\'enon--Heiles family
\[
H_4=\frac{1}{2}\left(p_1^2+p_2^2\right)
   +a q_1^4+b q_1^2q_2^2+c q_2^4.
\]
Although the mixed term $q_1^2q_2^2$ couples the two modes and removes Cartesian separability, the coefficient ratios
\[
a:b:c=1:2:1,\qquad 1:12:16,\qquad 1:6:1,\qquad 1:6:8
\]
are Liouville integrable; suitable inverse-square extensions remain integrable as well \cite{BakerEnolskiiFordy1995,ConteMusetteVerhoeven2005}. Thus, neither a quartic coupling nor the loss of the familiar oscillator integrals determines whether a hidden invariant survives. Hidden integrals can occur at isolated coefficient ratios, so the symmetric TTW deformation requires a coefficient-specific nonintegrability test.  Hamiltonian~\eqref{eq:intro-H} also interpolates between two important limits. For $\kappa=0$ and $\gamma\neq 0$, it reduces to the maximally superintegrable $k=1$ TTW, or Smorodinsky--Winternitz, system \cite{Evans1991,TremblayTurbinerWinternitz2009}}.  For $\gamma=0$ and $\kappa \neq 0$, it becomes the quartically coupled oscillator, or Contopoulos-type Hamiltonian, a standard model for resonance-driven Hamiltonian chaos \cite{ContopoulosPapadakiPolymilis1994, ContopoulosHarssoulaVoglisDvorak1999, ContopoulosEfthymiopoulosGiorgilli2003,AtkinsEzra1995, JimenezLaraLlibre2011}.
{The quartic product $X^2Y^2$ is the standard coupling of the Contopoulos oscillator and also appears in homogeneous two-mode reductions of Yang--Mills dynamics \cite{QuirozJuarezZuritaOlivaresPilonEscobarRuiz2026}. In the present setting it couples radial degrees of freedom carrying fixed centrifugal charges. The parameters $\kappa$ and $\gamma$ can then be varied independently: $\kappa$ sets the nonlinear exchange between the modes, whereas $\gamma$ fixes the geometry and connectivity of the accessible region.}

The inverse-square terms play a role beyond ordinary confinement.  For $\gamma>0$, the axes $X=0$ and $Y=0$ become impenetrable singular barriers, partitioning configuration space into invariant quadrants. These barriers do not restore integrability once $\kappa\neq0$, but they modify the transport geometry of the phase space.  In this sense, the system may be viewed as a Contopoulos oscillator dressed by TTW singular walls: the quartic term creates nonlinear resonances and chaotic layers, while the singular barriers restrict transport and attenuate the resulting chaos.  This places the model within the broader setting of near-integrable Hamiltonian dynamics, where resonances, invariant tori, separatrix layers, geometric instability, and chaotic transport organize the departure from regular motion \cite{Arnold1989,LichtenbergLieberman1992,Kandrup1997}.

Related perturbative mechanisms have been studied in singular and anisotropic natural Hamiltonians, where weak deformations can generate periodic-orbit continuation, nonintegrability obstructions, and chaotic dynamics \cite{DiacuSantoprete2001,Santoprete2002}. Recent studies of perturbed Hamiltonian and near-Hamiltonian planar systems have also addressed limit-cycle bifurcations produced by higher-order polynomial perturbations and by Melnikov-function mechanisms \cite{WangCao2025,LiuLiuLiCai2025}. Closely related (1:1) resonant Hamiltonian settings, including symmetric quartic perturbations and degenerate averaging constructions for periodic solutions, provide
a further point of contact with the present approach \cite{PerezRothenVidal2023,Vidal2026}.

Here we analyze this competition by combining perturbative nonintegrability theory with numerical phase-space diagnostics. After reducing the model to its dimensionless two-parameter form, we determine the stationary structure of the effective potential and the associated confinement or escape thresholds. The resulting Hamiltonian family connects the maximally superintegrable \(k=1\) TTW, or Smorodinsky--Winternitz, system with the quartically coupled Contopoulos oscillator, while retaining the singular inverse-square barriers that partition configuration space into invariant sectors.

Our main analytic result concerns the weak-coupling regime
\[
\kappa=\varepsilon\,a,\qquad a\neq 0,\qquad 0<|\varepsilon|\ll 1 .
\]
On each energy surface \({\cal H}=h>4\sqrt{\gamma}\), first-order resonant averaging yields a nondegenerate phase-locked periodic orbit near the symmetric resonance \(r=s=h/2\). The Poincaré map reduced to the fixed energy surface along this orbit has no unit characteristic multiplier. Hence, by Poincaré's nonintegrability criterion, there is no additional independent $C^1$ first integral on any invariant domain containing the orbit
\cite{Poincare1892,Kozlov1983,JimenezLaraLlibre2011,EscobarRuizJimenezLaraLlibreZurita2025,VerhulstBakri2024}. Since averaging is used only to construct the persistent periodic orbit in the near-integrable regime \cite{SandersVerhulstMurdock2007,BuicaLlibre2004}, this obstruction is local in phase space and perturbative in \(\kappa\).

Beyond the perturbative regime, we use Poincaré sections and finite-time Lyapunov heat maps to follow the transition from regular motion to mixed dynamics, using standard Hamiltonian-chaos diagnostics \cite{BenettinGalganiGiorgilliStrelcyn1980a,BenettinGalganiGiorgilliStrelcyn1980b,LichtenbergLieberman1992,DaSilvaMancheinBeimsAltmann2015}.
The use of finite-time Lyapunov diagnostics is part of a broader numerical approach to chaos detection in nonlinear dynamical systems, alongside alignment-index methods such as GALI \cite{MogesManosRacoveanuSkokos2025}. The maps in the $(\kappa,E)$ plane quantify the growth of chaotic response with energy and quartic coupling, while comparison with the barrier-free Contopoulos limit isolates the role of the TTW walls. The quartic interaction breaks the superintegrable organization through resonant coupling; the inverse-square barriers restrict phase-space transport and attenuate, but do not eliminate, the resulting chaotic response. Therefore, the model separates two effects that are often intertwined: integrability breaking by nonlinear resonance and chaos suppression by singular confinement.

\section{Generalities}

Let us consider, in the Euclidean space $\mathbb R^2$, the classical  Hamiltonian
\begin{equation}
H =
\frac{p_x^2+p_y^2}{2\,m}
+ m\,\omega^2(x^2+y^2)
+ \frac{g}{x^2}
+ \frac{g}{y^2}
+ A\, x^2 y^2 ,
\label{eq:Hdim}
\end{equation}
where \(m>0\) is the mass, \(\omega>0\) is the oscillator frequency, and \(g\) and \(A\) are real coupling constants. Here $(x,y)$ denote the Cartesian coordinates with canonical momentum variables $(p_x,p_y)$, respectively.  The singular-barrier regime corresponds to \(g > 0\), while the barrier-free Contopoulos limit is obtained by setting \(g=0\). The first two terms describe an isotropic harmonic oscillator in the plane, the inverse-square terms generate singular barriers along the coordinate axes, and the quartic interaction couples the two directions nonlinearly. For $A=0$ the system reduces to the $k=1$ TTW (equivalently, Smorodinsky--Winternitz) model, whereas the quartic deformation $A\,x^2y^2$ preserves the discrete reflection and exchange symmetries
\[
x\to -x,\qquad y\to -y,\qquad x\leftrightarrow y,
\]
but, as discussed below, breaks the integrable structure of the undeformed problem. For $g=0$, the Hamiltonian~\eqref{eq:Hdim} reduces to the isotropic oscillator with quartic coupling $A x^2y^2$, namely the quartically perturbed oscillator often referred to as the Contopoulos Hamiltonian.

It is useful to remove the inessential dimensional constants by introducing oscillator-adapted dimensionless variables. Since \(\omega\) has dimensions of inverse time, an energy scale is obtained from \(\omega\) only after choosing an action scale. We therefore introduce an arbitrary constant reference action \(J_0\) and define
\[
x=\sqrt{\frac{J_0}{m\omega}}\,X,
\qquad
y=\sqrt{\frac{J_0}{m\omega}}\,Y,
\]
\[
p_x=\sqrt{m\omega J_0}\,P_X,
\qquad
p_y=\sqrt{m\omega J_0}\,P_Y.
\]

We also introduce the dimensionless time
\[
\tau=\omega t,
\]
and the dimensionless Hamiltonian
\[
\mathcal{H}=\frac{H}{J_0\omega}.
\]

With these definitions, the extended Poincaré one-form transforms as
\[
p_x\,dx+p_y\,dy-H\,dt
=
J_0\left(
P_X\,dX+P_Y\,dY-\mathcal{H}\,d\tau
\right).
\]
Thus, after division by the constant factor \(J_0\), the variables \(X,Y,P_X,P_Y\) are canonical with respect to the dimensionless time \(\tau\). Equivalently, the equations of motion become
\[
\frac{dX}{d\tau}
=
\frac{\partial \mathcal{H}}{\partial P_X},
\qquad
\frac{dP_X}{d\tau}
=
-\frac{\partial \mathcal{H}}{\partial X},
\]
\[
\frac{dY}{d\tau}
=
\frac{\partial \mathcal{H}}{\partial P_Y},
\qquad
\frac{dP_Y}{d\tau}
=
-\frac{\partial \mathcal{H}}{\partial Y}.
\]

Substitution into the dimensional Hamiltonian gives
\begin{equation}
\mathcal{H}
=
\frac{1}{2}\left(P_X^2+P_Y^2\right)
+X^2+Y^2
+\frac{\gamma}{X^2}
+\frac{\gamma}{Y^2}
+\kappa \,X^2Y^2 ,
\label{eq:Hdimensionless}
\end{equation}
where
\[
\gamma=\frac{gm}{J_0^2},
\qquad
\kappa=\frac{A J_0}{m^2\omega^3}.
\]

Hamiltonian~\eqref{eq:Hdimensionless} shows that the classical dynamics depend only on the two dimensionless couplings
$(\gamma,\kappa)$. Thus, all trajectories related by a simultaneous change of the original dimensional parameters that leave $(\gamma,\kappa)$ fixed are dynamically equivalent up to the units set by $m,\omega$ and $J_0$.  Since \(J_0\) is an arbitrary reference action introduced only to define dimensionless variables, it is not an additional physical parameter. Equivalently, after nondimensionalization all actions may be measured in units of \(J_0\). In this convention the numerical value of \(J_0\) is set equal to one, and the dimensionless couplings are written as
\[
\gamma = g\,m,
\qquad
\kappa = \frac{A}{m^2\omega^3}.
\]
Several qualitative features are immediate. The singular terms partition configuration space by the barriers $X=0$ and $Y=0$ whenever $\gamma>0$, while the quartic term is globally confining for $\kappa>0$ and
destabilizing along suitable directions for $\kappa<0$. In addition, $\mathcal{H}$ is invariant under the discrete dihedral-type symmetries generated by
\[
(X,Y)\mapsto(-X,Y);\
(X,Y)\mapsto(X,-Y);\
(X,Y)\mapsto(Y,X),
\]
so that the phase-space structure may be analyzed within a single symmetry sector and then extended to the full system by reflection and exchange.

In what follows~\eqref{eq:Hdimensionless} will serve as the basic starting point for both the analytic characterization of the stationary structure and the numerical exploration of the emergence of nonintegrable dynamics.

\section{Superintegrable case: $\kappa=0$}

For $\kappa=0$ the Hamiltonian~(\ref{eq:Hdimensionless}) reduces to
\begin{equation}
\mathcal H_0 =
\frac{1}{2}(P_X^2+P_Y^2)
+ X^2 + Y^2
+ \frac{\gamma}{X^2}
+ \frac{\gamma}{Y^2},
\label{eq:H0}
\end{equation}
which corresponds to the symmetric $k=1$ TTW (Smorodinsky--Winternitz) system. In particular, the Hamiltonian~(\ref{eq:H0}) is separable in Cartesian coordinates,
\begin{equation}
\mathcal H_0 = H_X + H_Y, \nonumber
\end{equation}
with
\begin{equation}
H_X = \frac{1}{2}P_X^2 + X^2 + \frac{\gamma}{X^2},
\qquad
H_Y = \frac{1}{2}P_Y^2 + Y^2 + \frac{\gamma}{Y^2}. \nonumber
\end{equation}
{For phase-space functions $F$ and $G$, we use the canonical Poisson bracket
\[
\{F,G\}
=\frac{\partial F}{\partial X}\frac{\partial G}{\partial P_X}
-\frac{\partial F}{\partial P_X}\frac{\partial G}{\partial X}
+\frac{\partial F}{\partial Y}\frac{\partial G}{\partial P_Y}
-\frac{\partial F}{\partial P_Y}\frac{\partial G}{\partial Y}.
\]
The above quantities $H_X$ and $H_Y$ are individually conserved and obey
\begin{equation}
\{H_X,H_Y\}=0, \nonumber
\end{equation}
so the system is Liouville integrable.
}

In addition, the Hamiltonian admits a third independent quadratic (in momentum variables) integral of motion $I$ associated with separability in polar coordinates,
\begin{equation}
I =
L^2
+ 2\gamma\left(
\frac{Y^2}{X^2}
+
\frac{X^2}{Y^2}
\right),
\qquad
L = X P_Y - Y P_X, \nonumber
\label{eq:hidden} 
\end{equation}
which satisfies
\begin{equation}
\{ I , \mathcal H_0 \} = 0. \nonumber
\end{equation}
The three functionally independent constants of motion $(\mathcal H_0, H_X, I)$ render the system maximally superintegrable in two dimensions.

As a consequence, the phase space is foliated by invariant tori and all bounded classical trajectories are closed \cite{nekhoroshev1972action}. The quartic deformation ($\kappa\neq0$) breaks the integrability, it produces nonlinear resonances and chaotic dynamics (see below).

Fixing $H_X=E_X$, the associated equation of motion is
\begin{equation}
\dot X^2 = 2E_X - 2X^2 - \frac{2\gamma}{X^2}. \nonumber
\end{equation}
{Introducing $s_X=X^2$ gives
\[
\dot{s}_X^{\,2}=8E_Xs_X-8s_X^2-8\gamma.
\]
The Hamilton equations also imply the second-order equation
\[
\ddot s_X+8s_X=4E_X.
\]
Hence every non-equilibrium bounded solution can be written as
\[
X^2(t)=\frac{E_X}{2}+A_X\cos\!\left[2\sqrt{2}\,(t-t_{0X})\right],
\qquad
A_X=\frac{1}{2}\sqrt{E_X^2-4\gamma}.
\]
The expression with the opposite sign of the cosine describes the same orbit after the phase shift $t_{0X}\mapsto t_{0X}+\pi/(2\sqrt{2})$. The two constant roots $E_X/2\pm A_X$ of the squared first-order equation are turning-point values, not constant solutions of the original Hamilton equations, except at the threshold $E_X=2\sqrt{\gamma}$. At that threshold $A_X=0$ and $X^2=\sqrt{\gamma}$ is the equilibrium.

The same reasoning gives
\[
Y^2(t)=\frac{E_Y}{2}+A_Y\cos\!\left[2\sqrt{2}\,(t-t_{0Y})\right],
\qquad
A_Y=\frac{1}{2}\sqrt{E_Y^2-4\gamma},
\]
with the equilibrium $Y^2=\sqrt{\gamma}$ at $E_Y=2\sqrt{\gamma}$.}

\medskip
\noindent{\bf Properties.}
The motion is periodic with a common frequency
\begin{equation}
\omega_{\rm eff}=2\,\sqrt2 , \nonumber
\end{equation}
which is energy independent. The turning points are
\begin{equation}
X_{\pm}^2=\frac{E_X}{2}\pm A_X,
\qquad
Y_{\pm}^2=\frac{E_Y}{2}\pm A_Y, \nonumber
\end{equation}
requiring $E_X^2 \ge 4\gamma$ and $E_Y^2 \ge 4\gamma$ for bounded motion.

Since both degrees of freedom oscillate with identical frequency, the full trajectory $(X(t),Y(t))$ is periodic and all bounded orbits are closed. This explicit isochronous structure underlies the maximal superintegrability of the $\kappa=0$ system.

\section{Integrability breaking in the TTW
system: averaging theory and Poincar\'e obstruction}
\label{aversec}

For $\kappa \neq 0$, we now prove a local weak-coupling obstruction to the existence of an additional first integral for the quartically deformed \(k=1\) TTW system. The argument is perturbative in the quartic coupling and is based on
first-order resonant averaging near the symmetric \(1:1\) phase-locked orbit of the undeformed system. The averaged reduced return map has a simple fixed point whose corresponding Poincar\'e map has no unit
characteristic multiplier. By Poincar\'e's nonintegrability criterion, this excludes an additional independent \(C^1\) first integral on any invariant neighbourhood of that orbit.

The result is deliberately local in phase space and perturbative in \(\kappa\). It should not be interpreted as a global finite-coupling nonintegrability theorem for any \textit{large} $\kappa$.

\medskip
\noindent{\bf Proposition 1.}
{\textit{
Fix \(\gamma>0\), \(a\neq0\), and an energy
\[
{\cal H}=h>4\sqrt{\gamma}.
\]
In any fixed invariant quadrant of the configuration space, there exists \(\varepsilon_0>0\) such that, for
\[
0<|\varepsilon|<\varepsilon_0,
\qquad
\kappa=\varepsilon a,
\]
the quartically deformed Hamiltonian admits a periodic orbit
\(\Gamma_\varepsilon(h)\subset\{{\cal H}=h\}\) close to the unperturbed
phase-locked orbit
\[
r=s=\frac h2,
\qquad
\alpha=0 .
\]
For this orbit, the Poincaré map on the section $\theta=0$, restricted to the energy surface $H=h$ and expressed in the reduced coordinates $(r,\alpha)$, has no unit characteristic multiplier. 
}

Consequently, by the Poincar\'e nonintegrability criterion \cite{Poincare1892, LlibreValls2011, MeletlidouIchtiaroglou1994}, there is no second
\(C^1\) first integral defined on an invariant neighbourhood of
\(\Gamma_\varepsilon(h)\), whose differential is independent of
\(d{\cal H}\) along \(\Gamma_\varepsilon(h)\).
}

\begin{proof}
We work in the dimensionless normalization of
Eq.~\eqref{eq:Hdimensionless} and write
\[
\kappa=\varepsilon a,
\qquad
a\neq0,
\qquad
0<|\varepsilon|\ll1 .
\]
Thus
\begin{equation}
{\cal H}=
\frac12(P_X^2+P_Y^2)
+X^2+Y^2
+\frac{\gamma}{X^2}
+\frac{\gamma}{Y^2}
+\varepsilon aX^2Y^2 , \nonumber
\label{eq:avg-H}
\end{equation}
with \(\gamma>0\). The unperturbed Hamiltonian is
\begin{equation}
{\cal H}_0=
\left(
\frac12P_X^2+X^2+\frac{\gamma}{X^2}
\right)
+
\left(
\frac12P_Y^2+Y^2+\frac{\gamma}{Y^2}
\right). \nonumber
\label{eq:avg-H0}
\end{equation}
The equations of motion are
\begin{equation}
\dot X=P_X,
\qquad
\dot Y=P_Y, \nonumber
\end{equation}
and
\begin{equation}
\dot P_X=-2X+\frac{2\gamma}{X^3}
-2\varepsilon aXY^2,
\quad
\dot P_Y=-2Y+\frac{2\gamma}{Y^3}
-2\varepsilon aX^2Y .
\label{eq:avg-eq}
\end{equation}

For each one-dimensional singular oscillator, introduce non-canonical polar-type variables by
\begin{equation}
X^2=\frac r2+R(r)\cos\theta,
\qquad
P_X=-\frac{\sqrt2\,R(r)\sin\theta}{X}, \nonumber
\label{eq:avg-X-transform}
\end{equation}
and
\begin{equation}
Y^2=\frac s2+R(s)\cos(\theta+\alpha),
\qquad
P_Y=-\frac{\sqrt2\,R(s)\sin(\theta+\alpha)}{Y}, \nonumber
\label{eq:avg-Y-transform}
\end{equation}
where
\begin{equation}
R(q)=\frac12\sqrt{q^2-c^2},
\qquad
c=2\sqrt{\gamma}. \nonumber
\label{eq:avg-R}
\end{equation}
Since the barriers \(X=0\) and \(Y=0\) are impenetrable for \(\gamma>0\), these variables are well defined in any fixed invariant quadrant. Their admissible domain is
\begin{equation}
r>c,
\qquad
s>c . \nonumber
\label{eq:avg-domain}
\end{equation}
Direct substitution gives
\begin{equation}
\frac12P_X^2+X^2+\frac{\gamma}{X^2}=r,
\qquad
\frac12P_Y^2+Y^2+\frac{\gamma}{Y^2}=s. \nonumber
\end{equation}
Therefore
\begin{equation}
{\cal H}_0=r+s,
\qquad
\dot\theta=2\sqrt2,
\qquad
\dot\alpha=0. \nonumber
\label{eq:avg-unpert}
\end{equation}
Thus \(\theta\) is the fast angle and \(\alpha\) is the resonant phase difference.

In the variables \((r,\theta,s,\alpha)\), the Hamiltonian becomes
\begin{equation}
{\cal H}=
r+s
+\varepsilon a
\left(
\frac r2+R(r)\cos\theta
\right)
\left(
\frac s2+R(s)\cos(\theta+\alpha)
\right). \nonumber
\label{eq:avg-H-rtheta}
\end{equation}
Equations~\eqref{eq:avg-eq} imply
\begin{equation}
\dot r=
2\sqrt2\,\varepsilon a\,R(r)\sin\theta
\left(
\frac s2+R(s)\cos(\theta+\alpha)
\right), \nonumber
\label{eq:avg-rdot}
\end{equation}
and
\begin{equation}
\begin{aligned}
\dot\alpha
&=
\sqrt2\,\varepsilon a
\Bigg[
\left(
\frac r2+R(r)\cos\theta
\right)
\left(
1+\frac{s}{2R(s)}\cos(\theta+\alpha)
\right)
-
\left(
\frac s2+R(s)\cos(\theta+\alpha)
\right)
\left(
1+\frac{r}{2R(r)}\cos\theta
\right)
\Bigg]. \nonumber
\end{aligned}
\label{eq:avg-alphadot}
\end{equation}
Moreover,
\[
\dot\theta=2\sqrt2+O(\varepsilon).
\]
Hence, after restricting to a compact neighbourhood of the resonant point and choosing \(|\varepsilon|\) sufficiently small, we have \(\dot\theta>0\). Thus \(\theta\) can be used as the independent variable. With primes denoting differentiation with respect to \(\theta\),
\begin{equation}
r'=\varepsilon F_{11}(\theta,r,\alpha)+O(\varepsilon^2),
\qquad
\alpha'=\varepsilon F_{12}(\theta,r,\alpha)+O(\varepsilon^2). \nonumber
\label{eq:avg-normal-form}
\end{equation}

We now restrict to the energy surface
\begin{equation}
{\cal H}=h,
\qquad
h>2c=4\sqrt{\gamma}. \nonumber
\label{eq:avg-energy}
\end{equation}
Since
\[
\frac{\partial {\cal H}}{\partial s}=1+O(\varepsilon), 
\]
the energy equation can be solved locally for \(s\) by the implicit function theorem. In a compact neighbourhood of the resonant point,
\begin{equation}
\begin{aligned}
s
&=
h-r
-\varepsilon a
\left(
\frac r2+R(r)\cos\theta
\right)
\left(
\frac{h-r}{2}+R(h-r)\cos(\theta+\alpha)
\right)
+O(\varepsilon^2). \nonumber
\end{aligned}
\end{equation}
Consequently, at first order it is sufficient to set
\begin{equation}
s=h-r \ , \nonumber
\label{eq:avg-s-first-order}
\end{equation}
inside the \(O(\varepsilon)\) terms.

With this substitution, the first-order system is
\begin{equation}
F_{11}
=
aR(r)\sin\theta
\left(
\frac s2+R(s)\cos(\theta+\alpha)
\right), \nonumber
\label{eq:avg-F11}
\end{equation}
and
\begin{equation}
\begin{aligned}
F_{12}
&=
\frac a2
\Bigg[
\left(
\frac r2+R(r)\cos\theta
\right)
\left(
1+\frac{s}{2R(s)}\cos(\theta+\alpha)
\right)
-
\left(
\frac s2+R(s)\cos(\theta+\alpha)
\right)
\left(
1+\frac{r}{2R(r)}\cos\theta
\right)
\Bigg], \nonumber
\end{aligned}
\label{eq:avg-F12}
\end{equation}
where \(s=h-r\).

The unnormalized averaged first-order system is
\begin{equation}
f_1(r,\alpha)
=
\bigl(f_{11}(r,\alpha),f_{12}(r,\alpha)\bigr)
=
\int_0^{2\pi}
\bigl(F_{11},F_{12}\bigr)\,d\theta . \nonumber
\label{eq:avg-f1-def}
\end{equation}
Direct integration gives
\begin{equation}
f_{11}(r,\alpha)
=
-\pi aR(r)R(s)\sin\alpha , \nonumber
\label{eq:avg-f11}
\end{equation}
and
\begin{equation}
f_{12}(r,\alpha)
=
\frac{\pi a}{2}
\left[
(r-s)
+
\frac12
\left(
\frac{sR(r)}{R(s)}
-
\frac{rR(s)}{R(r)}
\right)
\cos\alpha
\right], \nonumber
\label{eq:avg-f12}
\end{equation}
again with \(s=h-r\).

We focus first on the in-phase branch \(\alpha=0\). On this branch, \(f_{11}=0\), and \(f_{12}=0\) is equivalent to
\begin{equation}
D(r)=0, \nonumber
\label{eq:avg-D-zero}
\end{equation}
where
\begin{equation}
D(r)
=
2(r-s)
+
\frac{sR(r)}{R(s)}
-
\frac{rR(s)}{R(r)},
\qquad
s=h-r. \nonumber
\label{eq:avg-D-def}
\end{equation}
The admissible interval is
\begin{equation}
c<r<h-c . \nonumber
\label{eq:avg-r-interval}
\end{equation}
{Set
\[
\widehat U(r)=r(h-r)+2R(r)R(h-r),
\qquad s=h-r.
\]
Since $ds/dr=-1$, differentiation along the fixed-energy surface gives
\[
\widehat U'(r)
=\left[s+\frac{rR(s)}{2R(r)}\right]
-\left[r+\frac{sR(r)}{2R(s)}\right],
\]
and therefore
\[
D(r)=-2\widehat U'(r).
\]
A further total derivative yields
\[
\widehat U''(r)
=-2-\frac{rs}{4R(r)R(s)}
-\frac{c^2}{8}
\left[
\frac{R(s)}{R(r)^3}+\frac{R(r)}{R(s)^3}
\right]<0,
\qquad c=2\sqrt{\gamma}.
\]
That way $D'(r)=-2\widehat U''(r)>0$ throughout the admissible interval.}
Moreover,
\[
\begin{aligned}
D(r) &\to -\infty
\quad\text{as}\quad r\to c^+,\\
D(r) &\to +\infty
\quad\text{as}\quad r\to (h-c)^- .
\end{aligned}
\]
Therefore, \(D\) has a unique zero in the admissible interval. By the exchange symmetry \(X\leftrightarrow Y\), this zero is
\begin{equation}
r^*=\frac h2,
\qquad
s^*=h-r^*=\frac h2 . \nonumber
\label{eq:avg-rstar-explicit}
\end{equation}
Thus
\begin{equation}
(r,\alpha)=(r^*,0) \nonumber
\label{eq:avg-zero-f1}
\end{equation}
is a zero of the averaged system.

This zero is simple. At \((r^*,0)\),
\begin{equation}
\frac{\partial f_{11}}{\partial r}=0,
\qquad
\frac{\partial f_{12}}{\partial \alpha}=0, \nonumber
\end{equation}
whereas
\begin{equation}
\frac{\partial f_{11}}{\partial \alpha}
=
-\pi aR(r^*)R(s^*),
\qquad
\frac{\partial f_{12}}{\partial r}
=
\frac{\pi a}{4}D'(r^*) . \nonumber
\end{equation}
Therefore
\begin{equation}
\det
\left(
\frac{\partial(f_{11},f_{12})}{\partial(r,\alpha)}
\right)_{(r^*,0)}
=
\frac{\pi^2a^2}{4}
R(r^*)R(s^*)D'(r^*) . \nonumber
\label{eq:avg-jacobian-det-pre}
\end{equation}
{At $r_*=s_*=h/2$,
\[
\widehat U''(r_*)
=-\frac{3h^2-16\gamma}{h^2-16\gamma},
\qquad
D'(r_*)=-2\widehat U''(r_*)
=2\frac{3h^2-16\gamma}{h^2-16\gamma}.
\]
Moreover,
\[
R(r_*)R(s_*)=\frac{h^2-16\gamma}{16}.
\]
Consequently,
\begin{equation}
\det\!\left[
\frac{\partial(f_{11},f_{12})}{\partial(r,\alpha)}
\right]_{(r_*,0)}
=\frac{\pi^2a^2}{4}R(r_*)R(s_*)D'(r_*)
=\frac{\pi^2a^2}{32}\left(3h^2-16\gamma\right)\neq0.
\label{eq:avg-jacobian-det}
\end{equation}
The nonvanishing follows from $h>4\sqrt{\gamma}$.}

The averaged system is \(2\pi\)-periodic in \(\theta\) and \(C^1\) on
compact subsets of the admissible strip \(c<r<h-c\). Since the averaged system $f_1$ has a simple zero at \((h/2,0)\), the first-order averaging theorem implies that, for all sufficiently small nonzero
\(|\varepsilon|\), the full \(\theta\)-system has a \(2\pi\)-periodic solution close to \((r^*,0)\). Equivalently, the original Hamiltonian flow has a periodic orbit \(\Gamma_\varepsilon(h)\) near
\begin{equation}
r=r^*,
\qquad
s=s^*,
\qquad
\alpha=0 . \nonumber
\label{eq:avg-unpert-periodic}
\end{equation}
Its period tends to
\begin{equation}
T_0=\frac{2\pi}{2\sqrt2}=\frac{\pi}{\sqrt2} \ , \nonumber
\label{eq:avg-period}
\end{equation}
as \(\varepsilon\to0\).

Let
\begin{equation}
R^*=R(r^*)=R(s^*) . \nonumber
\label{eq:avg-Rstar}
\end{equation}
The limiting unperturbed orbit is
\begin{equation}
\begin{aligned}
X^2(t)
&=
\frac{r^*}{2}
+
R^*\cos(2\sqrt2\,t+\theta_0),
\\
Y^2(t)
&=
\frac{s^*}{2}
+
R^*\cos(2\sqrt2\,t+\theta_0). \nonumber
\end{aligned}
\label{eq:avg-unpert-orbit}
\end{equation}
For \(\theta_0=0\) the corresponding initial condition is
\begin{equation}
X(0)=
\sigma_1
\sqrt{\frac{r^*}{2}+R^*},
\qquad
P_X(0)=0, \nonumber
\label{eq:avg-IC-unpert-X}
\end{equation}
\begin{equation}
Y(0)=
\sigma_2
\sqrt{\frac{s^*}{2}+R^*},
\qquad
P_Y(0)=0, \nonumber
\label{eq:avg-IC-unpert-Y}
\end{equation}
where
\begin{equation}
\sigma_1,\sigma_2\in\{+1,-1\}. \nonumber
\end{equation}
The signs label the invariant quadrants selected by the singular barriers.

For the perturbed orbit choose the phase so that \(\theta(0)=0\). Then
\begin{equation}
r_\varepsilon=r^*+O(\varepsilon),
\qquad
s_\varepsilon=s^*+O(\varepsilon),
\qquad
\alpha_\varepsilon=O(\varepsilon), \nonumber
\end{equation}
where \(s_\varepsilon\) is fixed by the exact energy constraint
\begin{equation}
h=
r_\varepsilon+s_\varepsilon
+
\varepsilon a
\left(
\frac{r_\varepsilon}{2}+R(r_\varepsilon)
\right)
\left(
\frac{s_\varepsilon}{2}
+
R(s_\varepsilon)\cos\alpha_\varepsilon
\right). \nonumber
\label{eq:avg-s-eps-energy}
\end{equation}
Thus
\begin{equation}
X(0;\varepsilon)
=
\sigma_1
\sqrt{
\frac{r_\varepsilon}{2}
+
R(r_\varepsilon)
},
\qquad
P_X(0;\varepsilon)=0, \nonumber
\label{eq:avg-IC-pert-X}
\end{equation}
and
\begin{equation}
Y(0;\varepsilon)
=
\sigma_2
\sqrt{
\frac{s_\varepsilon}{2}
+
R(s_\varepsilon)\cos\alpha_\varepsilon
}, \nonumber
\label{eq:avg-IC-pert-Y}
\end{equation}
\begin{equation}
P_Y(0;\varepsilon)
=
-\sigma_2
\frac{
\sqrt2\,R(s_\varepsilon)\sin\alpha_\varepsilon
}{
\sqrt{
\frac{s_\varepsilon}{2}
+
R(s_\varepsilon)\cos\alpha_\varepsilon
}
}. \nonumber
\label{eq:avg-IC-pert-PY}
\end{equation}
Equivalently,
\begin{equation}
\begin{aligned}
&\bigl(
X(0;\varepsilon),
P_X(0;\varepsilon),
Y(0;\varepsilon),
P_Y(0;\varepsilon)
\bigr)
 =
\left(
\sigma_1\sqrt{\frac{r^*}{2}+R^*},
0,
\sigma_2\sqrt{\frac{s^*}{2}+R^*},
0
\right)
+O(\varepsilon). \nonumber
\end{aligned}
\label{eq:avg-IC-asymptotic}
\end{equation}

Let \(P_\varepsilon\) be the Poincar\'e map on the local section \(\theta=0\), restricted to the energy surface \({\cal H}=h\). In the reduced coordinates \((r,\alpha)\),
\begin{equation}
P_\varepsilon(r,\alpha)-(r,\alpha)
=
\varepsilon f_1(r,\alpha)+O(\varepsilon^2). \nonumber
\label{eq:avg-Poincare-displacement}
\end{equation}
At the fixed point corresponding to the periodic orbit,
\[
(r_\varepsilon,\alpha_\varepsilon)\to(r^*,0),
\]
and hence
\begin{equation}
DP_\varepsilon(r_\varepsilon,\alpha_\varepsilon)
=
I+\varepsilon Df_1(r^*,0)+O(\varepsilon^2). \nonumber
\end{equation}
Therefore
\begin{equation}
\det
\bigl(
DP_\varepsilon(r_\varepsilon,\alpha_\varepsilon)-I
\bigr)
=
\varepsilon^2
\det
\left(
\frac{\partial(f_{11},f_{12})}{\partial(r,\alpha)}
\right)_{(r^*,0)}
+
O(\varepsilon^3). \nonumber
\label{eq:avg-multiplier-det}
\end{equation}
By Eq.~\eqref{eq:avg-jacobian-det} this determinant is nonzero for all sufficiently small nonzero \(\varepsilon\). Hence, the reduced Poincar\'e map has no unit characteristic multiplier at the constructed periodic
orbit.

Poincar\'e's nonintegrability criterion states that if a two-degree-of-freedom Hamiltonian system admits a second \(C^1\) first integral \(C\), with \(d{\cal H}\) and \(dC\) linearly independent along a periodic orbit, then the corresponding reduced Poincar\'e map must have a unit characteristic multiplier at the associated fixed point. The orbit constructed above violates this necessary condition. Thus, no such additional \(C^1\) first integral can exist on an invariant neighbourhood of \(\Gamma_\varepsilon(h)\). \end{proof}

\vspace{0.5cm}

\begin{Remark}

{At the anti-phase zero $(r,\alpha)=(h/2,\pi)$,
\[
\det\!\left[
\frac{\partial(f_{11},f_{12})}{\partial(r,\alpha)}
\right]_{(h/2,\pi)}
=-\frac{\pi^2a^2}{32}\left(h^2-48\gamma\right).
\]
Hence this zero is simple except at $h=4\sqrt{3\gamma}$.} This additional branch is not needed for the proposition above, but it shows that the resonant averaged dynamics has a second phase-locked family.
\end{Remark}

The proposition is local and perturbative. It proves an obstruction to an additional independent \(C^1\) first integral near the constructed phase-locked orbit for sufficiently small nonzero \(\kappa\). The finite-\(\kappa\) transition to mixed and chaotic dynamics is addressed below through Poincar\'e sections and Lyapunov diagnostics.

It is important to distinguish the periodic orbit constructed above from the local small-amplitude periodic orbits that emanate from equilibria of the effective potential. The orbit in Proposition 1 is obtained in the weak-coupling limit ($\kappa=\varepsilon a$), with $\varepsilon\to0$, at fixed energy $h>4\sqrt{\gamma}$. Therefore, it converges to a nontrivial phase-locked periodic orbit of the undeformed superintegrable system, not to a stationary point. By contrast, the equilibrium families discussed in Appendix C arise by taking a small-amplitude limit at fixed ($\kappa$), and collapse to a critical point of the potential. These are therefore different limiting mechanisms, although they become related in the additional low-energy limit ($h\to4\sqrt{\gamma}^{+}$) when ($\kappa=0$).

As a numerical consistency check, we also integrate a finite-\(\kappa\) corrected initial condition for a representative weak-coupling case. The result, reported in Appendix B, confirms the expected periodic motion on
the invariant diagonal branch and shows that the relative energy error remains at the level of numerical precision.

\section{Hamiltonian in diagonal coordinates}

To analyze the dynamics away from Cartesian separability, it is convenient to introduce the rotated variables
\begin{equation}
u = \frac{X+Y}{\sqrt{2}}, 
\qquad 
v = \frac{X-Y}{\sqrt{2}}, \nonumber
\end{equation}
with conjugate momenta
\begin{equation}
P_u = \frac{P_X+P_Y}{\sqrt{2}}, 
\qquad 
P_v = \frac{P_X-P_Y}{\sqrt{2}}. \nonumber
\end{equation}
The transformation is canonical because
\begin{equation}
P_X dX + P_Y dY = P_u du + P_v dv . \nonumber
\end{equation}

In these variables the kinetic and quadratic terms become
\begin{equation}
P_X^2+P_Y^2=P_u^2 + P_v^2,
\qquad
X^2+Y^2 = u^2+v^2. \nonumber
\end{equation}
The quartic interaction simplifies to
\begin{equation}
X^2 Y^2 = \frac{(u^2-v^2)^2}{4}. \nonumber
\end{equation}

The inverse-square contribution can be combined into a single rational form,
\begin{equation}
\frac{\gamma}{X^2}+\frac{\gamma}{Y^2}=2\gamma\left(\frac{1}{(u+v)^2}+\frac{1}{(u-v)^2}\right)=\frac{4\gamma (u^2+v^2)}{(u^2-v^2)^2}. \nonumber
\end{equation}

The Hamiltonian therefore takes the compact form
\begin{equation}
\mathcal H =\frac{1}{2}\left(P_u^2 + P_v^2\right)+u^2+v^2+\frac{4\gamma(u^2+v^2)}{(u^2-v^2)^2}+\frac{\kappa}{4}(u^2-v^2)^2.
\label{eq:Huv}
\end{equation}

The Hamiltonian~(\ref{eq:Huv}) makes the symmetry structure transparent. The singular barriers lie at $u=\pm v$ (corresponding to $X=0$ or $Y=0$), while the quartic interaction depends only on the invariant combination
$(u^2-v^2)^2$. In these coordinates, the dynamics along the diagonal ($v=0$) and anti-diagonal ($u=0$) directions can be analyzed directly, which is particularly useful for stability and Poincar\'e section studies.

\section{Critical Energies}

\begin{figure}[h]
\centering
\includegraphics[width=0.95\textwidth]{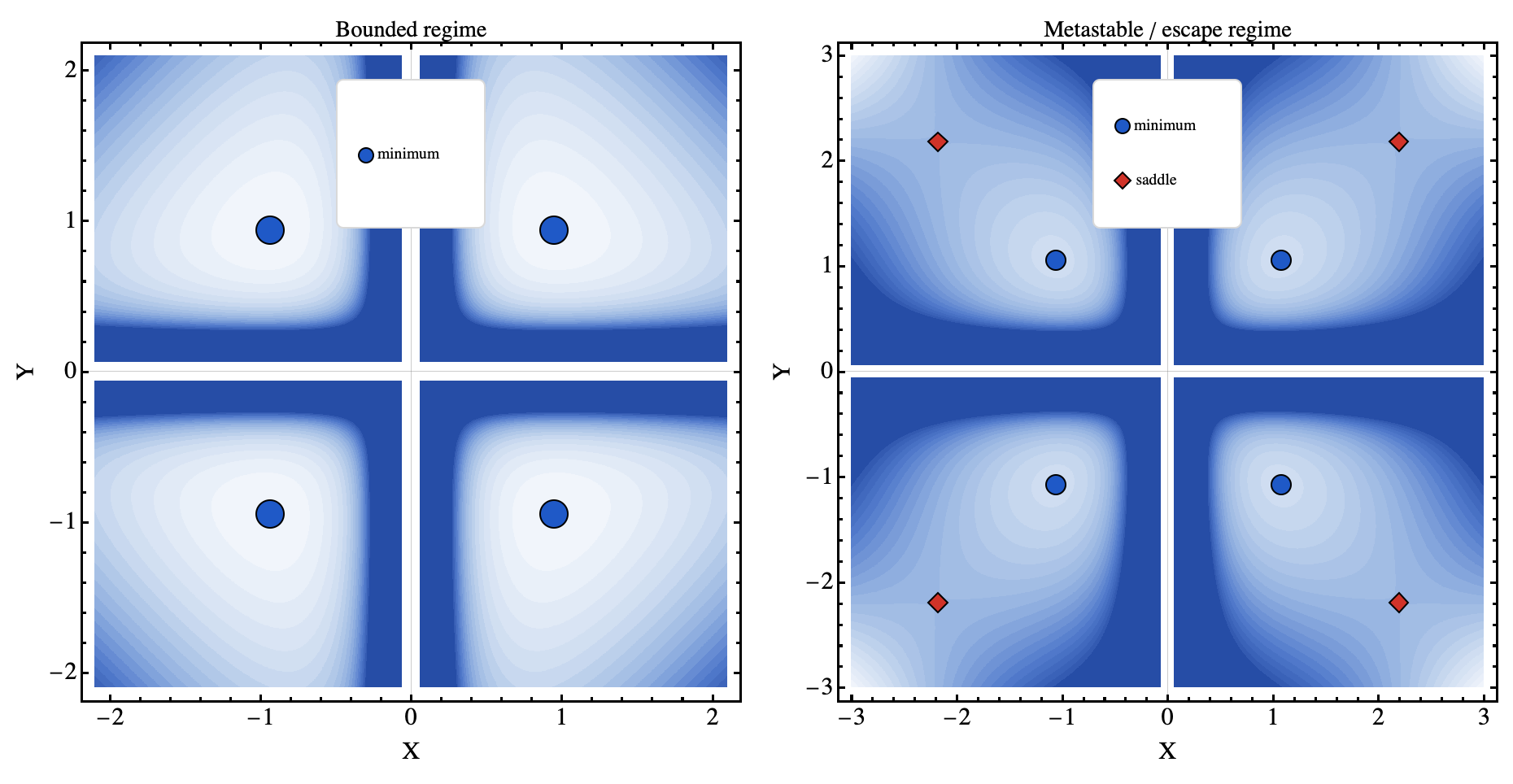}
\caption{
Effective-potential contours of (\ref{Vefff}) for $\gamma=1$.  Left: bounded regime, $\kappa=0.30$, with $E_{\min}\simeq4.2648$.  Right: metastable regime, $\kappa=-0.20$, with $E_{\min}\simeq3.7745$ and $E_c\simeq5.4087$.  Blue circles mark minima and red diamonds mark saddles.
}
\label{fig:potential_contours}
\end{figure}

\begin{figure}[h]
\centering
\includegraphics[width=0.98\textwidth]{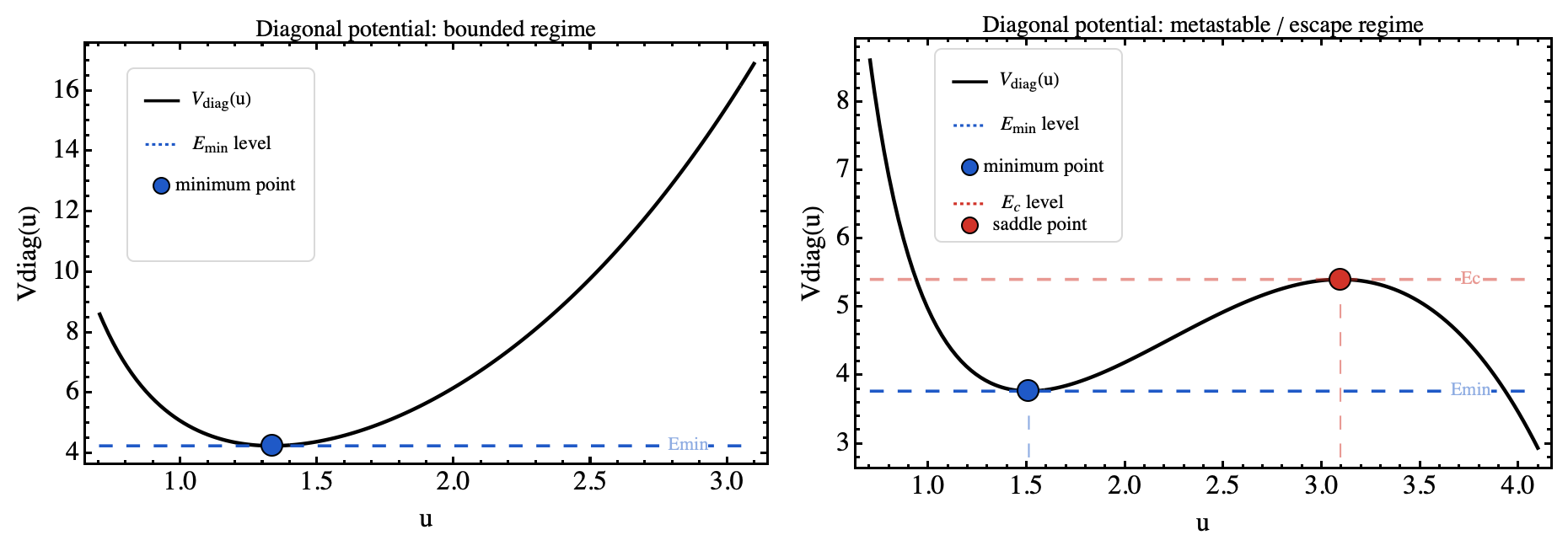}
\caption{
Diagonal restriction $V_{\mathrm{diag}}(u)=u^2+4\gamma/u^2+\kappa u^4/4$ for $\gamma=1$.  Left: bounded regime, $\kappa=0.30$.  Right: metastable regime, $\kappa=-0.20$.  Dashed lines indicate the critical energies, and markers denote stationary points.
}
\label{fig:diagonal_potential}
\end{figure}

The topology of the energy surface is controlled by the stationary points of the effective potential
\begin{equation}
V(u,v)=u^2+v^2+\frac{4\gamma (u^2+v^2)}{(u^2-v^2)^2}+\frac{\kappa}{4}(u^2-v^2)^2 . 
\label{efpot}
\end{equation}
For $\gamma>0$, it is convenient to characterize the stationary points in the original Cartesian variables
\[
X=\frac{u+v}{\sqrt{2}},\qquad
Y=\frac{u-v}{\sqrt{2}},
\]
for which
\begin{equation}
V(X,Y)=X^2+Y^2+\frac{\gamma}{X^2}+\frac{\gamma}{Y^2}+\kappa X^2Y^2 \ .
\label{Vefff}
\end{equation}
The corresponding contours are shown in Fig.~\ref{fig:potential_contours} for representative bounded
and metastable regimes. Because of the inverse-square barriers, any stationary point must satisfy $X\neq 0$ and $Y\neq 0$. Writing
\[
a=X^2>0,\qquad b=Y^2>0,
\]
the conditions $\partial_XV=\partial_YV=0$ become
\[
1-\frac{\gamma}{a^2}+\kappa b=0,
\qquad
1-\frac{\gamma}{b^2}+\kappa a=0.
\]
Subtracting these equations yields
\[
(a-b)\left(\kappa-\frac{\gamma(a+b)}{a^2b^2}\right)=0.
\]
The second factor is incompatible with the stationary equations: indeed, substituting
\[
\kappa=\frac{\gamma(a+b)}{a^2b^2}
\]
into the first equation gives
\[
a^2b+\gamma a=0,
\]
which is impossible for $a,b>0$. Therefore one must have $a=b$, i.e.
\[
X^2=Y^2.
\]
Hence all stationary points lie on the symmetry-related lines $X=\pm Y$, equivalently $v=0$ or $u=0$. It is therefore sufficient to analyze one representative branch, say $v=0$, where the potential reduces to the one-dimensional form
\begin{equation}
V_{\mathrm{diag}}(u)=
u^2+\frac{4\gamma}{u^2}+\frac{\kappa}{4}u^4 . \nonumber
\label{eq:Vdiag}
\end{equation}
Representative diagonal profiles of \(V_{\rm diag}\) are shown in Fig.~\ref{fig:diagonal_potential}, illustrating the unique confining minimum for \(\kappa>0\) and the minimum--saddle structure for
\(\kappa<0\).

Stationary points satisfy
\begin{equation}
\frac{dV_{\mathrm{diag}}}{du}
=
2u-\frac{8\gamma}{u^3}+\kappa u^3
=0, \nonumber
\end{equation}
which yields the cubic equation
\begin{equation}
\kappa z^3+2z^2-8\gamma=0,
\qquad
z=u^2. 
\label{eq:cubic}
\end{equation}

\subsection*{A. $\kappa>0$ (globally bounded case)}

For \(\gamma>0\) and \(\kappa>0\) the Eq.~(\ref{eq:cubic}) has a unique positive root \(z_0\). This root determines four symmetry-related minima of the full potential. On the representative branch \(v=0\), two of them are located
at \(u=\pm\sqrt{z_0}\); the other two lie on the branch \(u=0\). The minimum energy is
\[
E_{\min}
=
z_0+\frac{4\gamma}{z_0}+\frac{\kappa}{4}z_0^2 .
\]
Since $V(u,v)\to+\infty$ as $|(u,v)|\to\infty$ and also $V(u,v)\to+\infty$ as $u\to\pm v$ because of the inverse-square barriers, the energetically accessible region in configuration space is compact. Accordingly, for $E>E_{\min}$ the corresponding energy surface is compact, and no escape energy exists. Dynamical transitions at higher energies
therefore correspond to bifurcations of periodic orbits and changes in phase-space structure, rather than to a change in global topology.

\subsection*{B. $\kappa<0$ (unbounded case)}

For $\kappa<0$ the quartic term destabilizes the potential. Along the representative symmetry branch $v=0$,
\begin{equation}
V_{\mathrm{diag}}(u)\sim
-\frac{|\kappa|}{4}u^4
\qquad (u\to\infty),  \nonumber
\end{equation}
so the system is no longer globally bounded. {For $\kappa<0$, define
\[
f(z)=\kappa z^3+2z^2-8\gamma.
\]
Then
\[
f'(z)=3\kappa z^2+4z=z(3\kappa z+4).
\]
So $f(z)$ has a single positive critical point at
\[
z_*=-\frac{4}{3\kappa}=\frac{4}{3|\kappa|}\,,
\]}
which is a local maximum of $f$. Hence Eq.~(\ref{eq:cubic}) has two distinct positive roots if and only if $f(z_*)>0$, namely
\[
-\frac{2}{3\sqrt{3\gamma}}<\kappa<0.
\]
At the threshold
\[
\kappa=-\frac{2}{3\sqrt{3\gamma}},
\]
the two roots coalesce into a degenerate stationary point, whereas for
\[
\kappa<-\frac{2}{3\sqrt{3\gamma}}
\]
no trapping well exists.

When
\[
-\frac{2}{3\sqrt{3\gamma}}<\kappa<0,
\]
let $z_m<z_s$ denote the two positive roots of Eq.~(\ref{eq:cubic}). The smaller root $z_m$ gives a local minimum of $V_{\mathrm{diag}}$, while the larger root $z_s$ gives a saddle of the full two-dimensional potential. The associated critical (escape) energy is
\begin{equation}
E_c=z_s+\frac{4\gamma}{z_s}+\frac{\kappa}{4}z_s^2 . \nonumber
\end{equation}
For $E<E_c$ trajectories started in the well remain trapped in a bounded region of phase space. At $E=E_c$ a separatrix appears, and for $E>E_c$ trajectories can escape to infinity along the symmetry directions $v=0$ or $u=0$, equivalently $X=\pm Y$. Thus, in the destabilized regime
$\kappa<0$, the system exhibits a genuine topological transition of the energy surface.

\begin{figure}[h]
\centering
\includegraphics[width=0.98\textwidth]{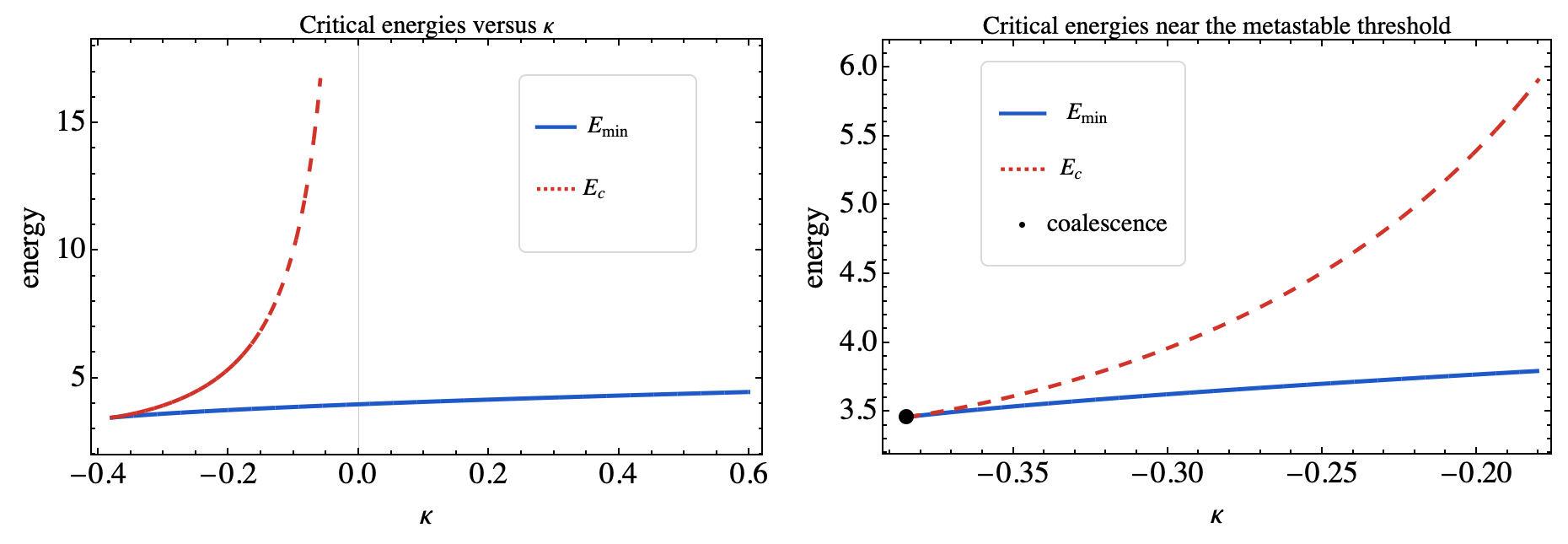}
\caption{
Critical energies as functions of $\kappa$ for $\gamma=1$.  The curves show
$E_{\min}$ and $E_c$, and the lower panel zooms near $\kappa_{\mathrm{th}}=-2/(3\sqrt{3\gamma})\simeq -0.3849$.  The black point
marks the coalescence $E_{\min}=E_c=2\sqrt{3}$.
}
\label{fig:critical_energies}
\end{figure}

The resulting critical energies as functions of \(\kappa\) are summarized
in Fig.~\ref{fig:critical_energies} for \(\gamma=1\).

\subsection*{C. Special free-barrier case $\gamma=0$}

When $\gamma=0$ the inverse-square barriers vanish and the potential (\ref{efpot}) simplifies to
\begin{equation}
V(u,v)=u^2+v^2+\frac{\kappa}{4}(u^2-v^2)^2 . \nonumber
\label{eq:Vgamma0}
\end{equation}
The origin $(u,v)=(0,0)$ is always a stationary point.

\medskip
\noindent{\bf 1. $\kappa>0$.}
In this case the quartic term is nonnegative. Since $V\to+\infty$ as $|(u,v)|\to\infty$, the origin is the unique global minimum,
\begin{equation}
E_{\min}=0, \nonumber
\end{equation}
and all energy surfaces are compact. No escape energy exists. The quadratic part dominates at low energy, and nonlinear effects appear only through amplitude-dependent frequency shifts and resonances.

\medskip
\noindent{\bf 2. $\kappa<0$.}
For negative $\kappa$ the quartic term destabilizes the system. Along the
symmetry branch $v=0$,
\begin{equation}
V_{\mathrm{diag}}(u)=u^2-\frac{|\kappa|}{4}u^4 . \nonumber
\end{equation}
Besides the minimum at $u=0$, two symmetric saddle points appear on this branch at
\begin{equation}
u_s^2=\frac{2}{|\kappa|}. \nonumber
\end{equation}
By symmetry there is an additional equivalent pair of saddles on the branch $u=0$. The corresponding critical (escape) energy is
\begin{equation}
E_c
=
V_{\mathrm{diag}}(u_s)
=
\frac{1}{|\kappa|}. \nonumber
\end{equation}

For $E<E_c$ motion is bounded near the origin. At $E=E_c$ a separatrix forms, and for $E>E_c$ trajectories can escape to infinity along the symmetry directions $v=0$ or $u=0$, equivalently $X=\pm Y$. Thus, for $\gamma=0$ and $\kappa<0$, the change at $E=E_c$ is a genuine topological transition of the energy surface.

\section{Numerical diagnostics in diagonal variables}

The numerical analysis is performed in the symmetry-adapted variables
\begin{equation}
u=\frac{X+Y}{\sqrt2},
\qquad
v=\frac{X-Y}{\sqrt2}, \nonumber
\end{equation}
with conjugate momenta
\begin{equation}
P_u=\frac{P_X+P_Y}{\sqrt2},
\qquad
P_v=\frac{P_X-P_Y}{\sqrt2}. \nonumber
\end{equation}
In these coordinates, the Hamiltonian is
\begin{equation}
{\cal H}
=
\frac12(P_u^2+P_v^2)
+u^2+v^2
+\frac{4\gamma(u^2+v^2)}{(u^2-v^2)^2}
+\frac{\kappa}{4}(u^2-v^2)^2 . \nonumber
\label{eq:Huv_RK}
\end{equation}
The singular walls are located at \(u=\pm v\), while the diagonal directions \(v=0\) and \(u=0\) correspond to the symmetry axes \(X=\pm Y\).

\subsection{ Flow and integration}
Writing
\begin{equation}
{\cal H}=\frac12(P_u^2+P_v^2)+V(u,v), \nonumber
\end{equation}
with
\begin{equation}
V(u,v)=
u^2+v^2
+\frac{4\gamma(u^2+v^2)}{(u^2-v^2)^2}
+\frac{\kappa}{4}(u^2-v^2)^2, \nonumber
\end{equation}
Hamilton's equations are
\begin{equation}
\dot u=P_u,\qquad
\dot v=P_v,\qquad
\dot P_u=-\partial_u V,\qquad
\dot P_v=-\partial_v V . \nonumber
\end{equation}
The derivatives entering the flow are
\begin{align}
\partial_u V
&=
2u
-\frac{8\gamma u(u^2+3v^2)}{(u^2-v^2)^3}
+\kappa u(u^2-v^2), \nonumber \\
\partial_v V 
&=
2v
+\frac{8\gamma v(3u^2+v^2)}{(u^2-v^2)^3}
-\kappa v(u^2-v^2). \nonumber
\end{align}
We integrate the first-order system
\[
\dot Z = F(Z),
\qquad
Z=(u,v,P_u,P_v),
\]
together with the associated variational equations using the adaptive fifth-order Tsitouras Runge--Kutta method, Tsit5, implemented in \texttt{DifferentialEquations.jl}. The relative and absolute tolerances are chosen so that the relative energy drift $\frac{|{\cal H}(t)-E|}{E}$  remains below the prescribed numerical threshold. The convergence of the Poincaré sections and finite-time Lyapunov exponents is checked by tightening the tolerances and increasing the final integration time.

\subsection{Poincar\'e sections}
\label{PSec}

\begin{figure}[t]
    \centering    \includegraphics[width=1.1\textwidth]{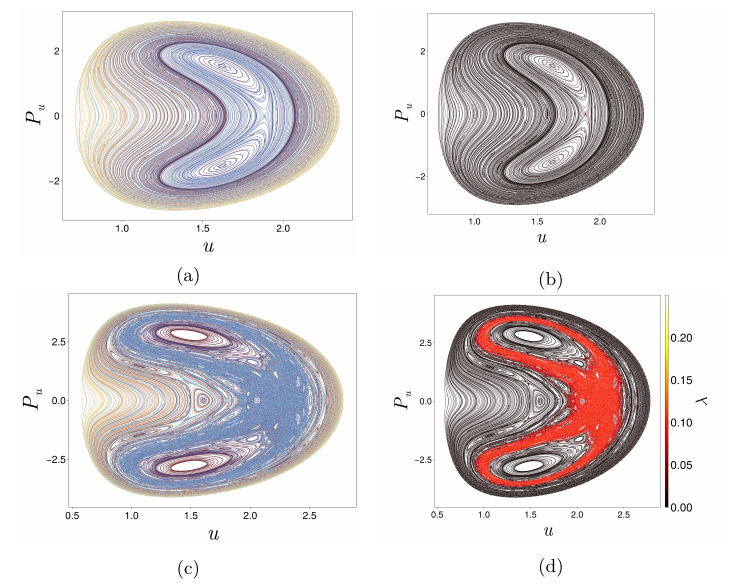}
    \caption{ Phase-space structure for \(\kappa=0.3\).  Each row shows a
    Poincar\'e section and the associated finite-time Lyapunov distribution.
    Top: \(E=2E_{\min}\), point \((III)\) in Fig.~\ref{fig:HeatMap}. Bottom:
    \(E=3E_{\min}\), point \((IV)\) in Fig.~\ref{fig:HeatMap}.}
    \label{fig:PoincareLyapunov}
\end{figure}

\begin{figure*}[]
    \centering
\includegraphics[width=0.79\textwidth]{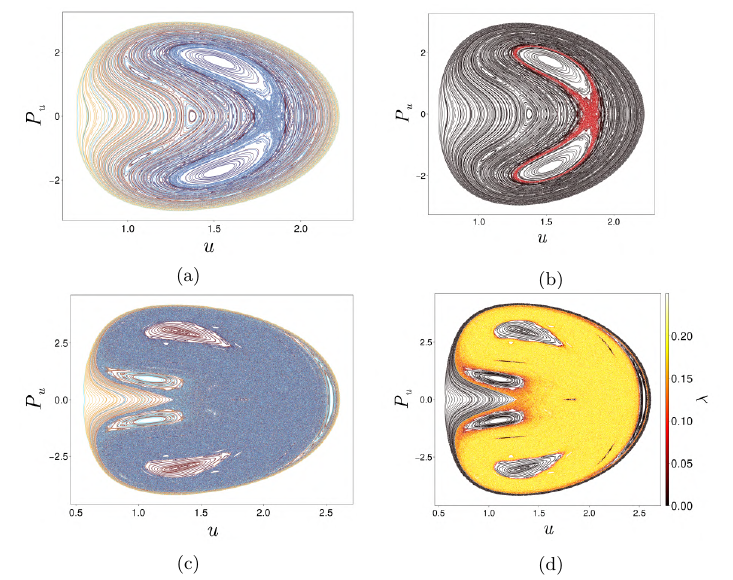}
    \caption{\small Phase-space structure for \(\kappa=0.5\).  Each row shows a
    Poincar\'e section and the associated finite-time Lyapunov distribution. Top: \(E=2E_{\min}\), point \((I)\) in Fig.~\ref{fig:HeatMap}. Bottom: \(E=3E_{\min}\), point \((II)\) in Fig.~\ref{fig:HeatMap}.}
    \label{fig:Fig5}
\end{figure*}

\begin{figure*}[]
    \centering
\includegraphics[width=0.7\textwidth]{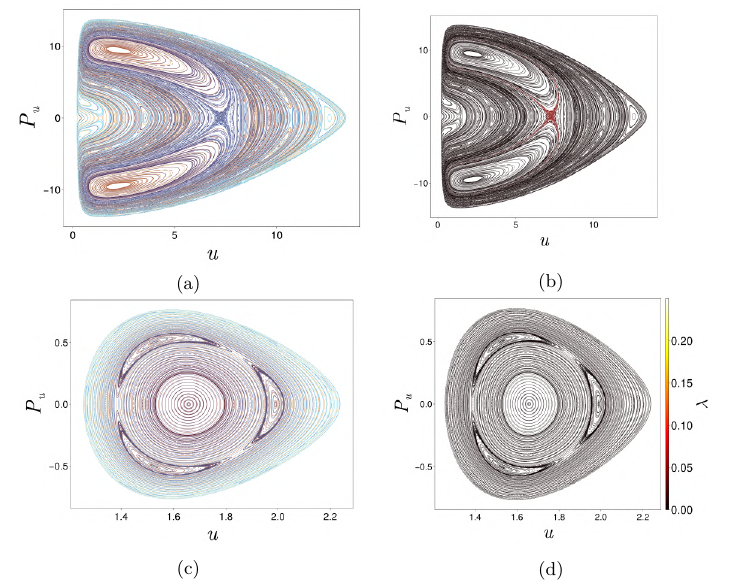}
    \caption{Near-escape dynamics in the metastable regime at \(E=0.99E_c\).  Each row shows a Poincar\'e section and the associated finite-time Lyapunov distribution. Top: \(\kappa=-0.01\). Bottom: \(\kappa=-0.3\).}
    \label{fig:Fig6}
\end{figure*}

For the representative Poincar\'e sections and their Lyapunov distributions, Figs. \ref{fig:PoincareLyapunov}, \ref{fig:Fig5}, \ref{fig:Fig6}, and \ref{fig:Fig8}, initial conditions are sampled on the two-dimensional energetically allowed domain of the section
\[
v=0,\qquad P_v>0 .
\]
The section coordinates are chosen as \((u,P_u)\). For fixed values of \((E,\gamma,\kappa)\), admissible initial points satisfy
\begin{equation}
\mathcal{D}_E
=
\left\{
(u,P_u):\;
2\bigl(E-V(u,0)\bigr)-P_u^2\ge 0
\right\}. \nonumber
\end{equation}
For each sampled pair \((u(0),P_u(0))\in \mathcal{D}_E\), we set
\begin{equation}
v(0)=0,\qquad
P_v(0)=
\sqrt{2\bigl(E-V(u(0),0)\bigr)-P_u(0)^2}, \nonumber
\end{equation}
so that the Hamiltonian constraint is satisfied and the crossing orientation is fixed by \(P_v(0)>0\).

The Poincar\'e section is therefore the oriented surface
\[
v=0,\qquad P_v>0,
\]
and the recorded coordinates are \((u,P_u)\). Crossings are obtained by linear interpolation between consecutive integration steps after discarding an initial transient. Each displayed section contains typically \(10^4\) intersections, collected from the ensemble of initial conditions in \(\mathcal{D}_E\).

The outer boundary of the accessible domain in the section is determined by
\[
P_v^2=2\bigl(E-V(u,0)\bigr)-P_u^2=0 .
\]

\subsection{Lyapunov exponents}
\label{secLE}

The maximal Lyapunov exponent is computed by integrating the Hamiltonian flow together with the variational equation
\[
\dot{\delta Z}=DF(Z)\,\delta Z,
\qquad
Z=(u,v,P_u,P_v).
\]
It is defined as
\[
\lambda_{\max}
=
\lim_{t\to\infty}
\frac{1}{t}
\ln
\frac{\|\delta Z(t)\|}{\|\delta Z(0)\|}.
\]
Numerically, we use the standard Benettin renormalization procedure. A deviation vector with initial norm one is evolved together with the reference trajectory and renormalized after fixed time intervals \(\tau\). If \(s_k\) denotes the stretching factor accumulated during the \(k\)-th interval, the finite-time estimate after \(N\) renormalizations is
\[
\lambda_{\max}(T)
=
\frac{1}{T}
\sum_{k=1}^{N}\ln s_k,
\qquad
T=N\tau .
\]
Convergence is monitored by increasing \(T\). Regular trajectories give \(\lambda_{\max}\simeq 0\), whereas chaotic trajectories give positive asymptotic values.

For the Lyapunov heat maps in the \((\kappa,E)\) plane, the ensemble of initial conditions is not sampled over the full two-dimensional Poincar\'e section. Instead, for each fixed pair \((\kappa,E)\), we use the one-parameter family
\[
v(0)=0,\qquad P_u(0)=0,\qquad P_v(0)>0,
\]
with \(u(0)\) varied over the admissible interval determined by
\[
2\bigl(E-V(u(0),0)\bigr)\ge 0 .
\]
For each admissible value of \(u(0)\), the momentum \(P_v(0)\) is fixed by the Hamiltonian constraint,
\[
P_v(0)
=
\sqrt{2\bigl(E-V(u(0),0)\bigr)} .
\]
The value displayed in each cell of the heat map is the ensemble average of the
finite-time maximal Lyapunov exponent over this prescribed one-parameter
sampling protocol. Thus, the heat maps should be interpreted as finite-time
Lyapunov diagnostics for a fixed family of initial conditions, rather than as
microcanonical averages over the full energy surface. They are not used as a
proof of finite-\(\kappa\) nonintegrability.

\medskip
\noindent{\bf Parameter scan and heat maps.}
For the representative Poincar\'e sections we fix \(\gamma=1\) and use
\[
\kappa\in\{0,\,0.01,\,0.05,\,0.1,\,0.3,\,0.5\},
\]
with
\[
E\in\{1.2E_{\min},\,1.5E_{\min},\,2E_{\min},\,3E_{\min}\}.
\]
Here \(E_{\min}\) is the minimum of the effective potential for the chosen \((\gamma,\kappa)\). In the metastable regime, when the saddle exists, we also consider
\[
E\in\{0.95E_c,\,0.97E_c,\,0.99E_c\}.
\]

For the Lyapunov heat maps, the same one-parameter sampling rule, integration time, renormalization interval, and numerical tolerances are used for the barrier-free Contopoulos case and for the singularly confined TTW deformation. The two heat maps are therefore compared under identical numerical conditions.

\begin{figure*}[t]
    \centering
    \includegraphics[width=1.0\textwidth]{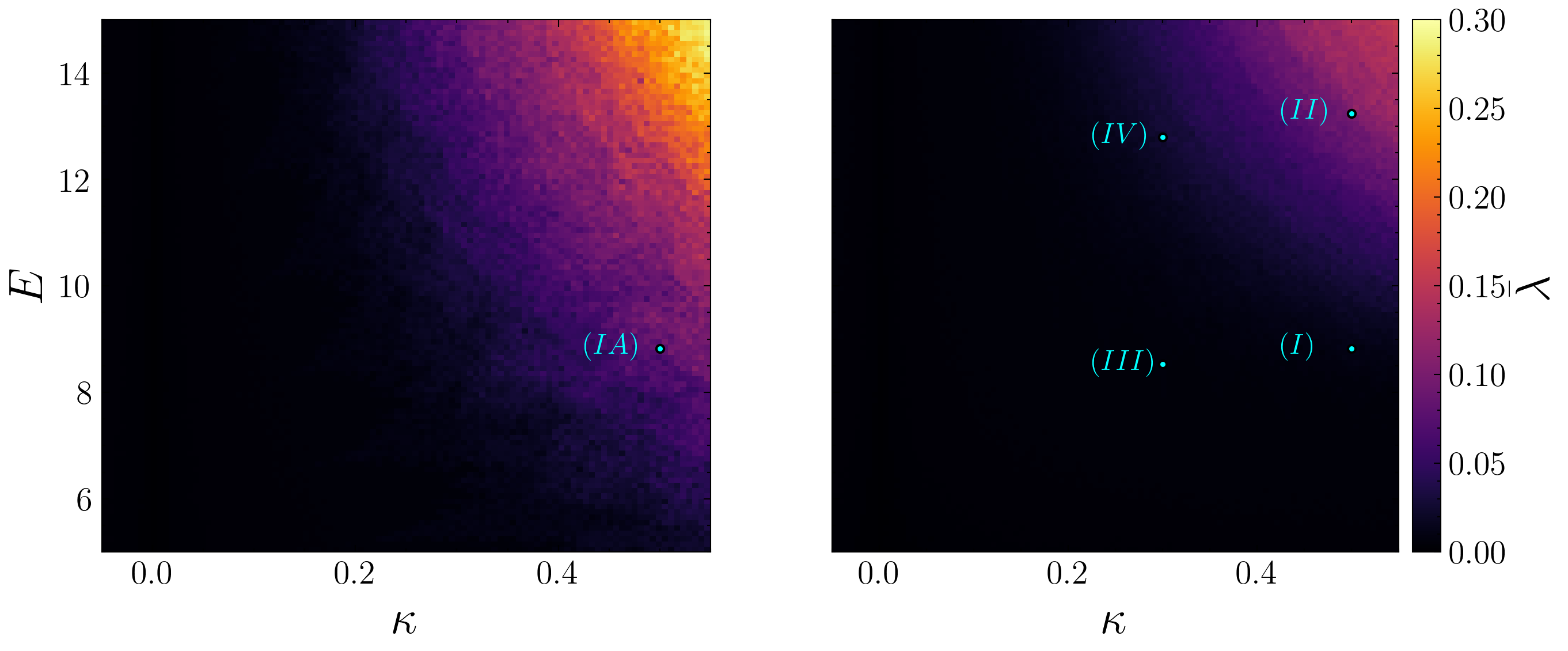}
    \caption{Finite-time Lyapunov maps for the prescribed one-parameter ensemble in the
\((\kappa,E)\) plane. The color scale gives the ensemble-averaged finite-time
maximal Lyapunov exponent computed from initial conditions satisfying
\(v(0)=0\), \(P_u(0)=0\), and \(P_v(0)>0\), with \(u(0)\) varied over the
admissible interval and \(P_v(0)\) fixed by the Hamiltonian constraint. Left:
barrier-free Contopoulos limit, \(\gamma=0\). Right: singularly confined TTW
deformation, \(\gamma=1\). Marked points correspond to the representative
sections in Figs.~\ref{fig:PoincareLyapunov}, \ref{fig:Fig5}, and
\ref{fig:Fig8}. The labels denote parameter values \((\kappa,E)\); the corresponding individual
representative Poincar\'e sections are computed separately using the
two-dimensional sampling protocol of Sec.~\ref{PSec}.}
    \label{fig:HeatMap}
\end{figure*}

\begin{figure*}[t]
    \centering
\includegraphics[width=1.0\textwidth]{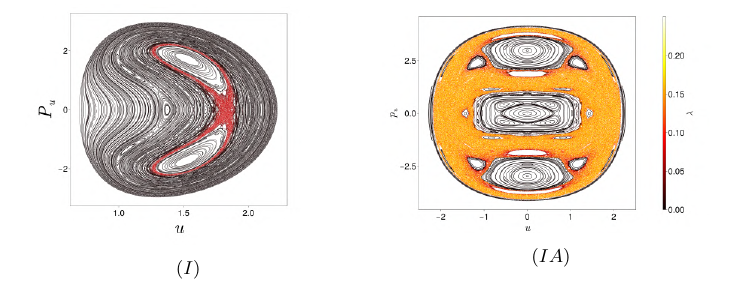}
    \caption{Representative comparison between the barrier-free and singularly confined systems at the marked points \((IA)\) and \((I)\) of Fig.~\ref{fig:HeatMap}.  The singular barriers suppress the chaotic response while preserving mixed phase-space structure.}
    \label{fig:Fig8}
\end{figure*}

\FloatBarrier

\section{Finite-time Lyapunov diagnostics and chaos attenuation}

The numerical diagnostics probe the finite-coupling regime beyond the
perturbative result. The averaging argument gives an obstruction to an
additional first integral near a phase-locked periodic orbit, while the
sections and Lyapunov maps reveal how mixed dynamics develops at larger
coupling: invariant curves break, chaotic layers grow, and the
accessible transport depends strongly on the singular barriers.

Figures~\ref{fig:PoincareLyapunov} and~\ref{fig:Fig5} show the progressive
breakup of invariant curves as either the energy or the quartic coupling is
increased.  For \(\kappa=0.3\) the section at \(E=2E_{\min}\) still contains
large regular domains, whereas increasing the energy to \(3E_{\min}\)
produces broader chaotic layers and a larger Lyapunov response.  The same
trend is amplified at \(\kappa=0.5\): the resonant web occupies a larger
fraction of the accessible section, and the Lyapunov distribution becomes
more intense.  Thus the quartic term acts as the driver of resonance overlap
and chaotic-layer growth.

The metastable regime in Fig.~\ref{fig:Fig6} shows a complementary effect.
Near the escape threshold, the phase-space geometry is strongly influenced by
the saddle structure of the effective potential.  For small negative
\(\kappa\) the accessible region is extended and the section develops broad
transport channels near the separatrix.  For larger negative \(\kappa\) the
well is tighter and the near-threshold dynamics remain more regular over the
displayed section.  This confirms that escape geometry and resonant coupling
jointly control the emergence of chaos in the unbounded regime.

The central finite-time ensemble diagnostic is Fig.~\ref{fig:HeatMap}.  In the barrier-free
Contopoulos limit, the ensemble-averaged finite-time Lyapunov exponent grows strongly with
\(\kappa\) and \(E\) under the prescribed sampling protocol, indicating that the quartic resonance opens efficient
transport channels across the accessible region.  With inverse-square TTW
barriers, the same qualitative trend persists, but the chaotic response is
substantially reduced.  The barriers partition configuration space into
invariant sectors and prevent transport across the coordinate axes.  They
therefore do not restore integrability; instead, they reduce the connectivity
and effective volume of the chaotic layers.

\begin{table}[htbp]
    \centering
    \begin{tabular}{c S[table-format=1.4] S[table-format=1.4] S[table-format=2.3]}
        \toprule
        Region in the $(\kappa,E)$ plane
        & {$\bar{\lambda}_{\rm 1D}$}
        & {$\bar{\lambda}_{\rm 2D}$}
        & {$\rho=\bar{\lambda}_{\rm 2D}/\bar{\lambda}_{\rm 1D}$} \\
        \midrule
        \textbf{I}   & 0.0062 & 0.0049 & 0.786 \\
        \textbf{IA}  & 0.0676 & 0.0841 & 1.245 \\
        \textbf{II}  & 0.0937 & 0.1241 & 1.325 \\
        \textbf{III} & 0.0018 & 0.0018 & 1.019 \\
        \textbf{IV}  & 0.0205 & 0.0217 & 1.060 \\
        \bottomrule
    \end{tabular}
    \caption{Finite-time Lyapunov response in the marked regions of Fig.~\ref{fig:HeatMap}. The entries are averages over the three ensemble sizes $N=150,200,250$. The one-dimensional protocol fixes $P_u(0)=0$, whereas the two-dimensional protocol samples the admissible section
    $\mathcal{D}_E=\{(u_0,P_{u,0}):P_{u,0}^2\leq 2[E-V(u_0,0)]\}$, with $v(0)=0$ and $P_v(0)>0$ fixed by the Hamiltonian constraint. The ratio $\rho$ measures the dependence of the finite-time Lyapunov average on the sampling protocol.}
    \label{tab:LLE_HeatMap_points}
\end{table}

Table~\ref{tab:LLE_HeatMap_points} compares the finite-time maximal Lyapunov response obtained from the one-dimensional heat-map ensemble with that obtained from the two-dimensional section ensemble. The ratios $\rho$ remain of order unity in all marked regions, showing that the qualitative trends are robust with respect to the sampling protocol. The two-dimensional sampling slightly lowers the response in region \textbf{I}, but increases it in the more chaotic regions \textbf{IA} and \textbf{II}, where sampling the full section reaches a larger portion of the resonant layer.

The suppression induced by the singular TTW barriers is quantified by comparing the barrier-free region \textbf{IA} with the confined region \textbf{I}. In the one-dimensional protocol,
\[
\frac{\bar{\lambda}_{\textbf{IA}}}{\bar{\lambda}_{\textbf{I}}}\simeq 10.8,
\]
whereas in the two-dimensional protocol,
\[
\frac{\bar{\lambda}_{\textbf{IA}}}{\bar{\lambda}_{\textbf{I}}}\simeq 17.2.
\]
Thus, the confined Lyapunov response is reduced to approximately $9.2\%$ of the barrier-free value in the one-dimensional ensemble and to approximately $5.8\%$ in the two-dimensional ensemble. This gives a quantitative measure of chaos attenuation by the inverse-square walls.

The positive confined values, however, show that this attenuation is not an integrability restoration. Region \textbf{II} gives the largest confined response, with $\bar{\lambda}_{\rm 1D}\simeq 0.0937$ and $\bar{\lambda}_{\rm 2D}\simeq 0.1241$, confirming that sufficiently large energy and quartic coupling still produce strong mixed dynamics. The singular barriers reduce phase-space transport and suppress the Lyapunov response, but the resonant quartic coupling remains dynamically effective.

Taken together Figs.~\ref{fig:PoincareLyapunov}--\ref{fig:Fig8} show that the deformed TTW system realizes chaos attenuation rather than integrability restoration.  The quartic interaction controls the strength of the resonant perturbation, while the inverse-square barriers control the geometry of accessible transport.  This separation between integrability breaking and transport suppression is the main numerical finding of the model. {This quantitative dynamical conclusion does not follow from the four-dimensional reduction, which identifies the symmetry origin of the centrifugal barriers but does not determine how they reorganize transport in the reduced nonintegrable flow.}

\section{Conclusion}

We have analyzed a symmetric quartic deformation of the \(k=1\) TTW system that interpolates between two limiting models: the maximally superintegrable Smorodinsky--Winternitz oscillator and the barrier-free quartically coupled Contopoulos oscillator. The deformation adds a resonant \(X^2Y^2\) interaction to two separable singular oscillators, whereas the inverse-square terms generate impenetrable barriers that split
configuration space into invariant sectors. This provides a minimal two-parameter setting in which resonant integrability breaking and singular geometric confinement can be studied simultaneously.

Our analytic result is local in phase space and perturbative in the quartic coupling. For \(\gamma>0\), and for sufficiently small nonzero \(\kappa=\varepsilon a\), first-order resonant averaging constructs a
nondegenerate phase-locked periodic orbit on every energy surface \({\cal H}=h>4\sqrt{\gamma}\). The reduced Poincaré map associated with this orbit has no unit characteristic multiplier. Poincaré's
nonintegrability criterion therefore rules out a second independent \(C^1\) first integral on any invariant neighbourhood containing the orbit. Thus, the quartic term gives a weak-coupling obstruction to Liouville--Arnold integrability, without implying a global finite-coupling nonintegrability theorem.

We also identified the local small-amplitude periodic families emanating from the equilibria of the effective potential; these Lyapunov-center-type orbits are distinct from the resonant phase-locked orbit used in the perturbative nonintegrability argument, since they collapse to stationary points rather than to nontrivial unperturbed TTW trajectories.

{The finite-coupling dynamics were probed numerically through Poincar\'e sections and finite-time Lyapunov maps. As the energy and quartic coupling increase, invariant curves break, resonant islands form, and chaotic layers grow. The central numerical question is not merely whether the barrier terms restore integrability, but what they do to the dynamics after integrability has been lost. The singular walls partition configuration space into invariant sectors,
restrict transport, and strongly attenuate the finite-time chaotic
response while leaving resonant islands and mixed phase-space structure
intact.

The four-dimensional reduction explains why the quartic deformation preserves the centrifugal charges while coupling the radial modes, but symmetry considerations alone do not rule out a hidden reduced integral. The periodic-orbit criterion supplies that obstruction in the weak-coupling regime. At finite coupling, the barriers act on a different level. Under the prescribed sampling protocols, the confined finite-time Lyapunov response is approximately $9.2\%$ of the barrier-free value in the one-dimensional ensemble and approximately $5.8\%$ in the two-dimensional section ensemble. These protocol-specific ratios quantify an order-of-magnitude attenuation of the chaotic response, while the positive confined response and the surviving mixed phase-space structure show that integrability is not restored. The system therefore realizes chaos attenuation by singular confinement rather than integrability restoration.}

\section{Declaration of competing interest}

The authors declare that they have no known competing financial interests or personal relationships that could have appeared to influence the work reported in this paper.

\section{Data availability}

No external datasets were used in this work. The trajectories are generated from analytic formulas and initial conditions reported in the main text.

\bigskip

\section{Acknowledgments}

A.M. Escobar Ruiz would like to thank the support from UAM research grant CBI-SA-391-26 PAPDI 2026. J. Llibre is partially supported by the Agencia Estatal de Investigaci\'on of Spain grant PID2022-136613NB-100.

\section*{Appendix A. Computational details of the numerical simulations}
\addcontentsline{toc}{section}{Appendix A. Computational Details of the Numerical Simulations}

The numerical simulations were performed by integrating the Hamiltonian equations
together with the associated tangent-space dynamics using the adaptive fifth-order
Runge--Kutta method Tsit5 implemented in \texttt{DifferentialEquations.jl}. The
representative Poincar\'e sections and their associated finite-time Lyapunov
distributions were computed from ensembles of initial conditions sampled on the
two-dimensional energetically allowed domain of the section \(v=0\), \(P_v>0\).
For fixed \((E,\gamma,\kappa)\), the section coordinates \((u,P_u)\) were sampled
inside
\[
\mathcal{D}_E
=
\left\{
(u,P_u):\;
2\bigl(E-V(u,0)\bigr)-P_u^2\ge 0
\right\},
\]
and \(P_v(0)\) was then fixed by the Hamiltonian constraint,
\[
P_v(0)
=
\sqrt{2\bigl(E-V(u(0),0)\bigr)-P_u(0)^2}.
\]
For each parameter set, a total of 675 independent trajectories were evolved.
Each trajectory was integrated up to a final simulation time \(t=5000\), which
was sufficient to obtain stable Poincar\'e sections and converged finite-time
estimates of the largest Lyapunov exponent for the representative cases shown.
The numerical integrations were carried out with relative and absolute
tolerances fixed at \(10^{-8}\).

For the Lyapunov heat maps, a systematic exploration of parameter space was
performed on a two-dimensional \(100\times100\) grid in the \((E,\kappa)\) plane,
covering the intervals \(E\in[4,15]\) and \(\kappa\in[-0.05,0.55]\). In this
case, the ensemble of initial conditions was not sampled over the full
two-dimensional Poincar\'e section. Instead, for each grid point, up to 150
dynamically admissible initial conditions were generated from the prescribed
one-parameter family
\[
v(0)=0,\qquad P_u(0)=0,\qquad P_v(0)>0,
\]
with \(u(0)\) varied over the admissible interval and \(P_v(0)\) determined by
\[
P_v(0)=\sqrt{2\bigl(E-V(u(0),0)\bigr)}.
\]
The Hamiltonian equations and the associated variational system were integrated
using the adaptive Tsit5 solver from \texttt{DifferentialEquations.jl}, with
relative and absolute tolerances fixed at \(10^{-6}\). Each trajectory was
evolved up to a final integration time \(t=1000\), and the largest Lyapunov
exponent was computed using the standard Benettin algorithm. The computations
over the parameter grid were parallelized using multithreading in Julia.

Consequently, the heat-map values should be interpreted as ensemble-averaged
finite-time Lyapunov diagnostics for this fixed one-parameter sampling protocol,
rather than as microcanonical averages over the full energy surface. The same
sampling rule and numerical parameters were used for the barrier-free
Contopoulos limit and for the singularly confined TTW deformation.

For the regional averages reported in Table~\ref{tab:LLE_HeatMap_points}, the marked regions of Fig.~\ref{fig:HeatMap} were recomputed using both the one-dimensional ensemble $P_u(0)=0$ and the two-dimensional admissible-section ensemble $\mathcal{D}_E$. For each protocol, the averages were evaluated with $N=150,200,250$ trajectories, and the values displayed in Table~\ref{tab:LLE_HeatMap_points} are the corresponding averages over these three ensemble sizes.

\section*{Appendix B. Numerical check of the phase-locked orbit}
\label{app:phase-locked-check}
\addcontentsline{toc}{section}{Appendix B. Numerical check of the phase-locked orbit}

We provide a numerical consistency check of the phase-locked orbit constructed by averaging. The dimensionless Hamiltonian is
\begin{equation}
{\cal H}=
\frac{1}{2}\left(P_X^2+P_Y^2\right)
+X^2+Y^2+\frac{\gamma}{X^2}
+\frac{\gamma}{Y^2}
+\kappa X^2Y^2 . \nonumber
\label{eq:app_H}
\end{equation}
We use
\begin{equation}
\gamma=1,\qquad \kappa=0.05,\qquad h=6 . \nonumber
\end{equation}
Since \(h>4\sqrt{\gamma}\), the averaged system has the phase-locked solution
\begin{equation}
r_\ast=s_\ast=\frac{h}{2},\qquad \alpha=0 . \nonumber
\end{equation}
At \(\kappa=0\) and \(\theta_0=0\),
\begin{equation}
X(0)^2=Y(0)^2=\frac{h}{4}+R_\ast,\qquad
P_X(0)=P_Y(0)=0, \nonumber
\end{equation}
where
\begin{equation}
R_\ast=
\frac{1}{2}
\sqrt{\left(\frac{h}{2}\right)^2-4\gamma}. \nonumber
\end{equation}
For the parameters above,
\begin{equation}
R_\ast=1.1180339887,\qquad
\frac{h}{4}+R_\ast=2.6180339887 . \nonumber
\end{equation}

For finite \(\kappa\) we impose the exact energy constraint on the diagonal branch
\begin{equation}
X^2=Y^2=z,\qquad P_X=P_Y=0 . \nonumber
\end{equation}
This gives
\begin{equation}
2z+\frac{2\gamma}{z}+\kappa z^2=h, \nonumber
\end{equation}
or
\begin{equation}
\kappa z^3+2z^2-hz+2\gamma=0 . \nonumber
\label{eq:app_cubic}
\end{equation}
The positive root connected to the unperturbed outer turning point is
\begin{equation}
z_0=2.4413875326 . \nonumber
\end{equation}
Thus
\begin{equation}
\begin{aligned}
X(0)&=Y(0)=1.5624940104, \nonumber \\
P_X(0)&=P_Y(0)=0 . \nonumber
\end{aligned}
\label{eq:app_ic}
\end{equation}
The initial value satisfies \(H=h\) to the displayed precision.

By exchange symmetry, this initial condition remains on the invariant
diagonal branch
\begin{equation}
X(t)=Y(t),\qquad P_X(t)=P_Y(t). \nonumber
\end{equation}
Therefore the rectilinear \((X,Y)\) projection is expected; the
periodicity is seen in the closed \((X,P_X)\) projection and in the
time traces. The equations were integrated with working precision \(70\),
accuracy goal \(25\), precision goal \(25\), and an eighth-order
explicit Runge--Kutta scheme over
\begin{equation}
T=20T_0,\qquad T_0=\frac{\pi}{\sqrt{2}}, \nonumber
\end{equation}
using \(2000\) steps per unperturbed period.

\begin{figure*}[h]
\centering
\includegraphics[width=0.99\linewidth]{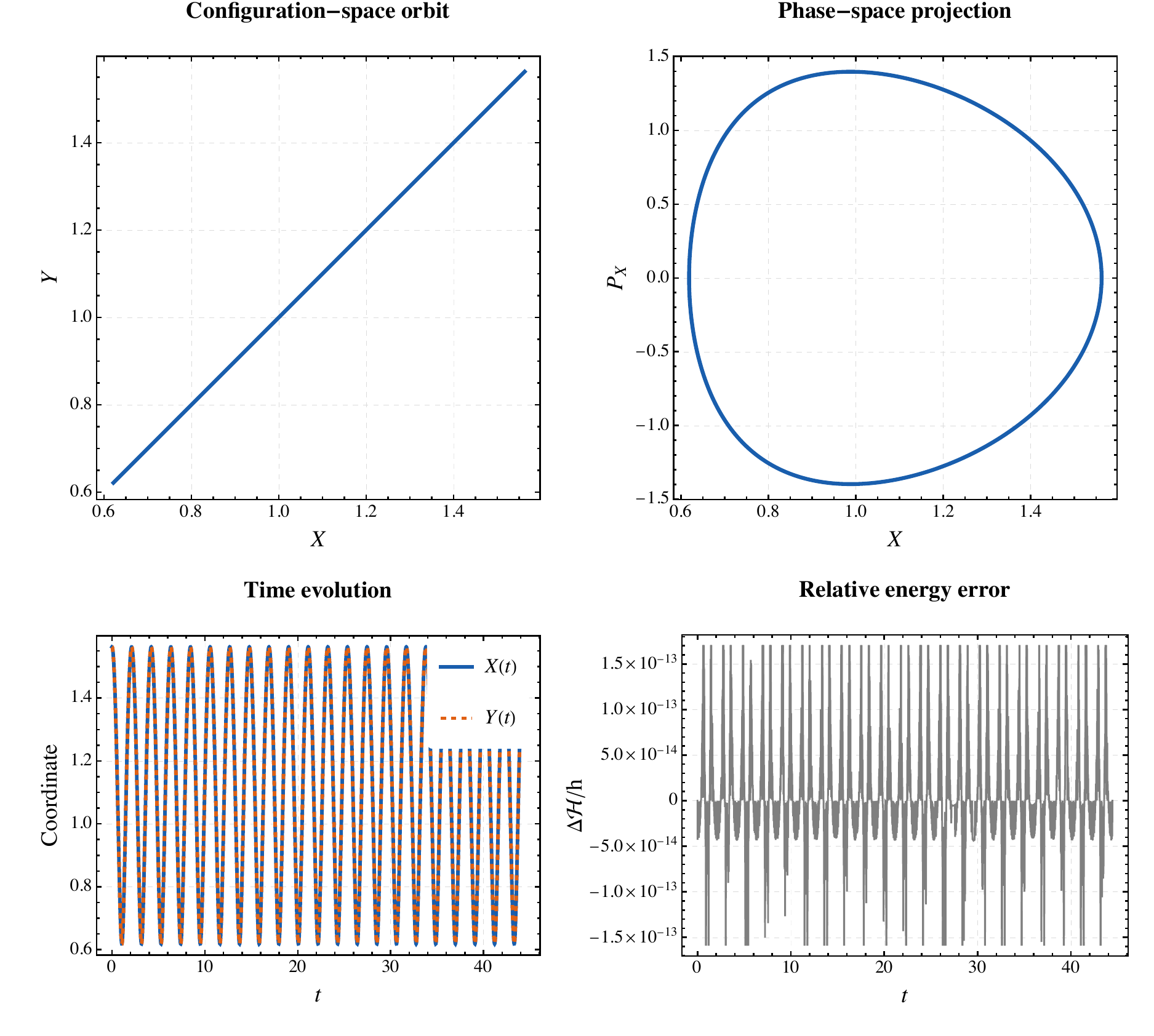}
\caption{
Numerical check of the phase-locked orbit for
\(\gamma=1\), \(\kappa=0.05\), and \(h=6\). The finite-\(\kappa\) corrected initial condition is \(X(0)=Y(0)=1.5624940104\), \(P_X(0)=P_Y(0)=0\). The straight \((X,Y)\) trace reflects the invariant diagonal branch \(X=Y\); the \((X,P_X)\) projection is closed, and the relative energy error remains of order \(10^{-13}\).
}
\label{fig:app_periodic_orbit}
\end{figure*}

Energy conservation was monitored through
\begin{equation}
\Delta {\cal H}(t)={\cal H}(t)-h,\qquad
\delta_{\cal H}(t)=\frac{\Delta {\cal H}(t)}{h}. \nonumber
\end{equation}
The sampled errors were
\begin{align}
\max |\Delta {\cal H}| &= 2.58\times 10^{-12},&
\max |\delta_{\cal H}| &= 4.31\times 10^{-13},\nonumber \\
\langle |\Delta {\cal H}| \rangle &= 2.65\times 10^{-13},&
\langle |\delta_{\cal H}| \rangle &= 4.43\times 10^{-14}. \nonumber
\end{align}
Thus the relative energy error remains of order \(10^{-13}\). The corresponding configuration-space orbit, phase-space projection, time series, and relative energy error are displayed in
Fig.~\ref{fig:app_periodic_orbit}.

For fixed energy, the averaging argument also shows that the
phase-locked orbit is isolated in the reduced Poincar\'e section in the
weak-coupling regime. The averaged zero \((r,\alpha)=(h/2,0)\) is simple, and the associated reduced return map satisfies
\begin{equation}
\det(DP_\kappa-I)\neq 0 \nonumber
\end{equation}
for sufficiently small nonzero \(\kappa\). Thus, the orbit is not part
of a continuous curve of fixed points of the reduced map, modulo the
trivial time parametrization and discrete symmetry-related copies. As
\(h\) varies, these orbits form a one-parameter family.

\FloatBarrier

\section*{Appendix C. Periodic orbits emanating from the equilibria}
\label{}
\addcontentsline{toc}{section}{Appendix C. Periodic orbits emanating from the equilibria}

We briefly record the small-amplitude periodic orbits which emanate from the
stationary points of the potential. This construction is distinct from the
phase-locked periodic orbit obtained in Sec.~4. The latter persists, for
\(\kappa=\varepsilon a\), from a nontrivial \(1:1\) resonant orbit of the
unperturbed system at fixed energy
\[
    H=h>4\sqrt{\gamma}.
\]
Thus, in the limit \(\varepsilon\to0\), the orbit of Sec.~4 converges to a periodic orbit of \(H_0\), not to an equilibrium. By contrast, the orbits considered here are local small-amplitude families whose amplitude tends to zero at fixed \(\kappa\), and therefore collapse to a stationary point.

We use the Cartesian form
\[
    H=
    \frac12(P_X^2+P_Y^2)
    +X^2+Y^2
    +\frac{\gamma}{X^2}
    +\frac{\gamma}{Y^2}
    +\kappa X^2Y^2,
    \qquad \gamma>0.
\]
The equilibria satisfy
\[
    P_X=P_Y=0,
    \qquad
    \partial_XV=\partial_YV=0.
\]
Since the singular barriers exclude \(X=0\) and \(Y=0\), all equilibria lie in the invariant quadrants. Writing
\[
    X_e^2=Y_e^2=z_e>0,
\]
one obtains
\[
    1-\frac{\gamma}{z_e^2}+\kappa z_e=0,
\]
or equivalently
\[
    \kappa z_e^3+z_e^2-\gamma=0.
\]
Thus the equilibria are
\[
    (X_e,Y_e,P_{X,e},P_{Y,e})
    =
    (\sigma_1\sqrt{z_e},\sigma_2\sqrt{z_e},0,0),
    \qquad
    \sigma_1,\sigma_2\in\{+1,-1\}.
\]

Let
\[
    \chi=\sigma_1\sigma_2.
\]
For the representative equilibrium in a fixed quadrant, introduce local coordinates
\[
    X=\sigma_1\sqrt{z_e}+\xi,
    \qquad
    Y=\sigma_2\sqrt{z_e}+\eta,
    \qquad
    P_X=p_\xi,
    \qquad
    P_Y=p_\eta .
\]
The Hessian of \(V\) at this equilibrium is
\[
    D^2V(X_e,Y_e)
    =
    \begin{pmatrix}
        8+8\kappa z_e & 4\kappa\chi z_e\\
        4\kappa\chi z_e & 8+8\kappa z_e
    \end{pmatrix}.
\]
Although the sign of the off-diagonal entries depends on the quadrant, the spectrum is independent of \(\sigma_1\) and \(\sigma_2\). Hence the squared linear normal frequencies are
\[
    \Omega_+^2=8+12\kappa z_e,
    \qquad
    \Omega_-^2=8+4\kappa z_e.
\]
The corresponding local configuration-space eigendirections are
\[
    e_+
    =
    \frac1{\sqrt2}(1,\chi),
    \qquad
    e_-
    =
    \frac1{\sqrt2}(1,-\chi).
\]

When
\[
    \Omega_+^2>0,
    \qquad
    \Omega_-^2>0,
\]
the equilibrium is elliptic at the linear level. If the corresponding Lyapunov-center nonresonance condition is satisfied, namely if the frequency of the selected normal mode is not an integer multiple of the other normal frequency, then each simple purely imaginary pair generates a local family of periodic orbits. If exactly one of the two squared frequencies is positive, the equilibrium is of saddle--center type. In that case, the simple imaginary pair still generates a local Lyapunov family under the usual saddle--center version of the Lyapunov-center nonresonance hypotheses.

The branch tangent to \(e_+\) is the local diagonal branch. At a turning point, its leading initial condition is
\[
    X(0;\rho)
    =
    \sigma_1\sqrt{z_e}
    +
    \frac{\rho}{\sqrt2}
    +
    \mathcal O(\rho^2),
\]
\[
    Y(0;\rho)
    =
    \sigma_2\sqrt{z_e}
    +
    \frac{\chi\rho}{\sqrt2}
    +
    \mathcal O(\rho^2),
\]
\[
    P_X(0;\rho)=P_Y(0;\rho)=0,
\]
and its period satisfies
\[
    T_+(\rho)
    =
    \frac{2\pi}{\Omega_+}
    +
    \mathcal O(\rho^2).
\]
Here \(\rho>0\) is the small amplitude parameter; any fixed normalization constant has been absorbed into \(\rho\).

The branch tangent to \(e_-\) is the local anti-diagonal branch. At a turning point, its leading initial condition is
\[
    X(0;\rho)
    =
    \sigma_1\sqrt{z_e}
    +
    \frac{\rho}{\sqrt2}
    +
    \mathcal O(\rho^2),
\]
\[
    Y(0;\rho)
    =
    \sigma_2\sqrt{z_e}
    -
    \frac{\chi\rho}{\sqrt2}
    +
    \mathcal O(\rho^2),
\]
\[
    P_X(0;\rho)=P_Y(0;\rho)=0,
\]
and
\[
    T_-(\rho)
    =
    \frac{2\pi}{\Omega_-}
    +
    \mathcal O(\rho^2).
\]

In both cases
\[
    H(\rho)=E_e+\mathcal O(\rho^2),
\]
where
\[
    E_e
    =
    2z_e+\frac{2\gamma}{z_e}+\kappa z_e^2.
\]
Moreover,
\[
    \lim_{\rho\to0}
    (X(0;\rho),Y(0;\rho),P_X(0;\rho),P_Y(0;\rho))
    =
    (X_e,Y_e,0,0).
\]
This is the precise sense in which these periodic orbits emanate from the equilibrium.

For completeness, we also express the leading initial conditions in the rotated variables
\[
    u=\frac{X+Y}{\sqrt2},
    \qquad
    v=\frac{X-Y}{\sqrt2},
    \qquad
    P_u=\frac{P_X+P_Y}{\sqrt2},
    \qquad
    P_v=\frac{P_X-P_Y}{\sqrt2}.
\]
The equilibrium becomes
\[
    u_e
    =
    \frac{\sigma_1+\sigma_2}{\sqrt2}\sqrt{z_e},
    \qquad
    v_e
    =
    \frac{\sigma_1-\sigma_2}{\sqrt2}\sqrt{z_e},
    \qquad
    P_{u,e}=P_{v,e}=0.
\]
For the representative first-quadrant equilibrium,
\[
    \sigma_1=\sigma_2=1,
    \qquad
    \chi=1,
\]
one has
\[
    (u_e,v_e,P_{u,e},P_{v,e})
    =
    (\sqrt{2z_e},0,0,0).
\]
In this quadrant, the \(e_+\)-branch becomes
\[
    u(0;\rho)
    =
    \sqrt{2z_e}
    +
    \rho
    +
    \mathcal O(\rho^2),
\]
\[
    v(0;\rho)=\mathcal O(\rho^2),
    \qquad
    P_u(0;\rho)=P_v(0;\rho)=0.
\]
The \(e_-\)-branch, at a turning point, becomes
\[
    u(0;\rho)
    =
    \sqrt{2z_e}
    +
    \mathcal O(\rho^2),
\]
\[
    v(0;\rho)
    =
    \rho
    +
    \mathcal O(\rho^2),
    \qquad
    P_u(0;\rho)=P_v(0;\rho)=0.
\]
Equivalently, on the Poincar\'e section
\[
    v=0,
    \qquad
    P_v>0,
\]
the same anti-diagonal branch can be represented by
\[
    u(0;\rho)
    =
    \sqrt{2z_e}
    +
    \mathcal O(\rho^2),
\]
\[
    v(0;\rho)=0,
    \qquad
    P_u(0;\rho)=0,
\]
\[
    P_v(0;\rho)
    =
    \Omega_-\rho
    +
    \mathcal O(\rho^2).
\]
The analogous expressions in the other invariant quadrants are obtained by using the signs \(\sigma_1,\sigma_2\) in the Cartesian formulae above, or equivalently by applying the corresponding reflections to the first-quadrant expressions.

Thus, the phase-locked orbit of Sec.~\ref{aversec} and the present equilibrium families arise from different limiting procedures. The former is obtained from
\[
    \kappa=\varepsilon a,
    \qquad
    \varepsilon\to0,
\]
at fixed
\[
    h>4\sqrt{\gamma},
\]
and converges to a nontrivial periodic orbit of the unperturbed system. The latter are obtained from
\[
    \rho\to0
\]
at fixed \(\kappa\), and collapse to a stationary point. They are connected only by the additional small-amplitude energy limit; for \(\kappa=0\), this is
\[
    h\to4\sqrt{\gamma}^{+}.
\]
This distinction is useful conceptually: the phase-locked orbit is a weak-coupling continuation at fixed energy, whereas the present local families are small-amplitude continuations from stationary points. They should therefore not be identified, except in the additional limiting process in which the energy of the unperturbed phase-locked orbit approaches the equilibrium energy.

\bibliographystyle{unsrt}
\bibliography{ref}

\end{document}